\documentclass[12pt]{article}
\pdfoutput=1

\usepackage{scicite}

\usepackage{times}
\usepackage{bm}
\usepackage{graphicx}
\usepackage{color}
\usepackage{amsmath}
\usepackage{amssymb}
\usepackage{array}
\usepackage{booktabs}

\newenvironment{sciabstract}{%
\begin{quote} \bf}
{\end{quote}}

\newcounter{lastnote}

\usepackage{graphicx} 

\usepackage{bm}
\usepackage{txfonts}

\newcommand{\sub}[1]{$_{\mathrm {#1}}$}
\newcommand{\subm}[1]{_{\mathrm {#1}}}
\newcommand{\sps}[1]{$^{\mathrm {#1}}$}
\newcommand{\spsm}[1]{^{\mathrm {#1}}}

\renewcommand{\deg}{^{\circ}}
\newcommand{\Tc}{T\subm{c}}

\newcommand{\Hcc}{H\subm{c2}}

\newcommand{\C}{C}

\newcommand{\kF}{\bm{k}\subm{F}}

\newcommand{\vnode}{\bm{v}\subm{F}\spsm{node}}
\newcommand{\vmin}{\bm{v}\subm{F}\spsm{min}}
\newcommand{\BS}{Bi$_2$Se$_3$}
\newcommand{\CBS}[1]{Cu$_{#1}$Bi$_2$Se$_3$}
\newcommand{\cbs}{Cu$_{x}$Bi$_2$Se$_3$}
\newcommand{\cbsbm}{Cu$_{\bm{x}}$Bi$_{\bm{2}}$Se$_{\bm{3}}$}
\newcommand{\bs}{Bi$_2$Se$_3$}

\title{\Large\bf 
Thermodynamic evidence \\for nematic superconductivity in Cu$_{\bm{x}}$Bi$_{\bm{2}}$Se$_{\bm{3}}$} 
\author{
Shingo~Yonezawa$^{1\ast}$, Kengo~Tajiri$^{1}$, Suguru Nakata$^{1}$, Yuki~Nagai$^{2}$, \\ Zhiwei~Wang$^{3,4}$, Kouji~Segawa$^{3,5}$,
Yoichi~Ando$^{3,4}$, Yoshiteru~Maeno$^1$\\
\\
\normalsize{$^{1}$Department of Physics, Graduate School of Science, Kyoto University,}\\
\normalsize{Kitashirakawa-Oiwake-Cho, Sakyo, Kyoto 606-8502, Japan}\\[0.2cm]
\normalsize{$^{2}$CCSE, Japan Atomic Energy Agency,}\\ \normalsize{178-4-4 Wakashiba, Kashiwa, Chiba 277-0871, Japan}\\[0.2cm]
\normalsize{$^{3}$The Institute of Scientific and Industrial Research, Osaka University,}\\
\normalsize{8-1 Mihogaoka, Ibaraki, Osaka 567-0047, Japan}\\[0.2cm]
\normalsize{$^{4}$Institute of Physics II, University of Cologne,}\\
\normalsize{77 Z\"{u}lpicher Str., K\"{o}ln 50937, Germany}\\[0.2cm]
\normalsize{$^{5}$Department of Physics, Graduate School of Science, Kyoto Sangyo University,}\\
\normalsize{Motoyama, Kamigamo, Kita, Kyoto 603-8555, Japan}\\[0.3cm]
\\
\normalsize{$^\ast$To whom correspondence should be addressed; }\\
\normalsize{E-mail: yonezawa@scphys.kyoto-u.ac.jp}
}

\date{\normalsize \it \today}

\begin{document} 


\baselineskip24pt
\sloppy


\maketitle 


\noindent
{\bf One-sentence summary:}\\
Spontaneous rotational symmetry breaking revealed in \cbs\ evidences novel odd-parity nematic superconductivity.

\clearpage


\begin{sciabstract} 
Unconventional superconductivity is characterized by the spontaneous symmetry breaking of the macroscopic superconducting wavefunction in addition to the gauge symmetry breaking, such as rotational-symmetry breaking with respect to the underlying crystal-lattice symmetry.
Particularly, superconductivity with spontaneous rotational-symmetry breaking in the wavefunction {\em amplitude} and thus in {\em bulk} properties, not yet reported previously, is intriguing and can be termed ``nematic'' superconductivity in analogy to nematic liquid-crystal phases.
Here, based on specific-heat measurements of the single-crystalline Cu$_{\bm{x}}$Bi$_{\bm{2}}$Se$_{\bm{3}}$ under accurate magnetic-field-direction control, we report thermodynamic evidence for nematic superconductivity, namely, clear two-fold-symmetric behavior in a trigonal lattice.
The results indicate realization of an ``odd-parity nematic'' state, feasible only by macroscopic quantum condensates and distinct from nematic states in liquid crystals.
The results also confirm topologically non-trivial superconductivity in Cu$_{\bm{x}}$Bi$_{\bm{2}}$Se$_{\bm{3}}$.
\end{sciabstract}

\begin{figure}
\begin{center}
\includegraphics[width=16cm]{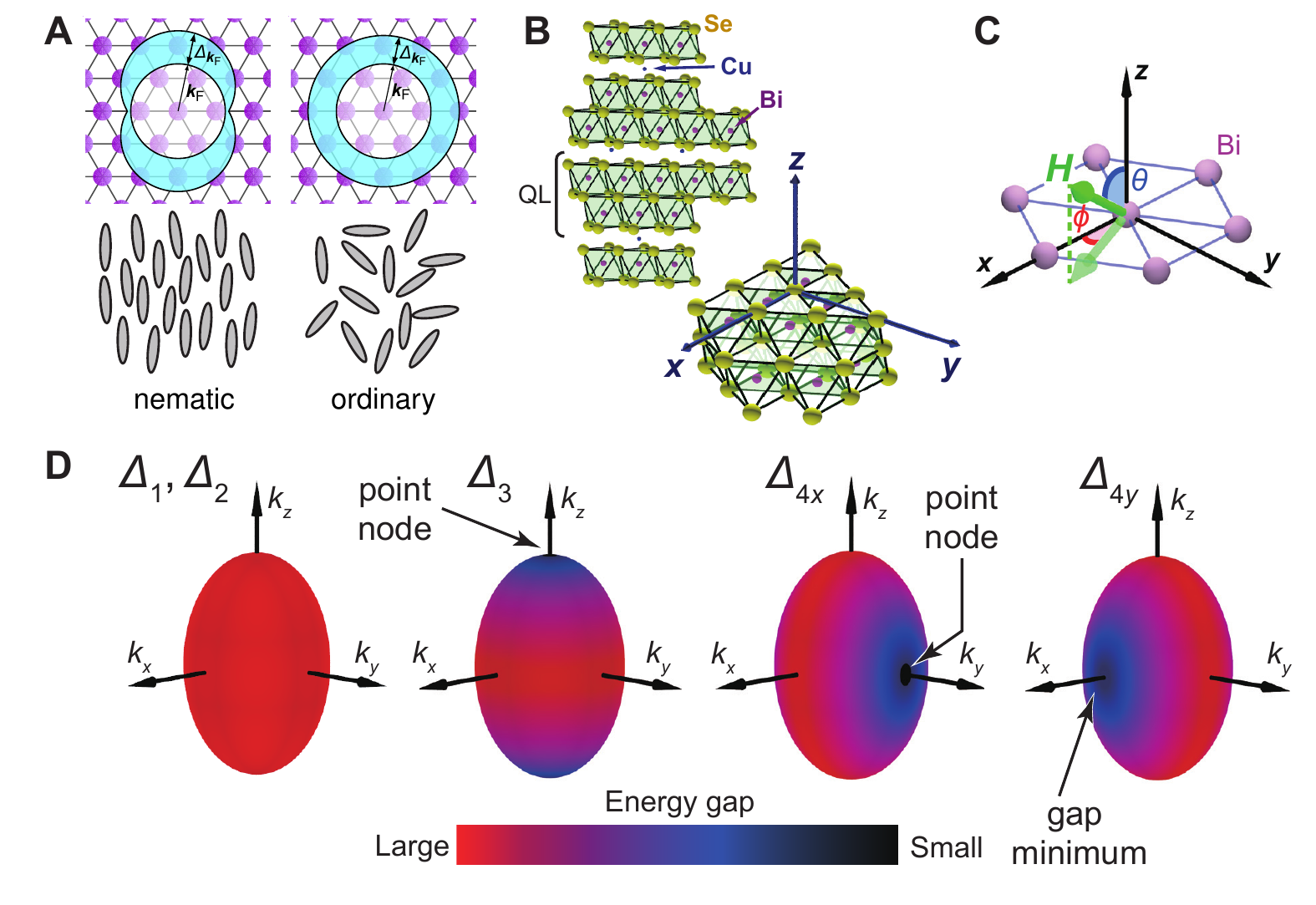}
\end{center}
\caption{
\baselineskip24pt
{\bf Candidate material for nematic superconductivity: \cbsbm.}
{\bf A}. Comparison of nematic and ordinary superconductivity in a hexagonal crystal system, with a nematic liquid-crystal phase and an ordinary liquid phase. The thickness of the blue region in the top panels illustrate the superconducting gap amplitude in the reciprocal space. The gray ovals in the bottom panels represent molecules in a liquid-crystal system.
{\bf B}. Crystal structure of \cbs\ with $x \sim 0.3$. The structure of the quintuple layer (QL) is shown in the right bottom figure.
{\bf C}. Definitions of the axes and field angles with respect to the crystal structure. The purple spheres represent Bi atoms. 
{\bf D}. Schematic description of SC gap structures $\Delta_1$, $\Delta_2$, $\Delta_3$, $\Delta_{4x}$, and $\Delta_{4y}$ proposed for \cbs\ in Refs.~\cite{Fu2010.PhysRevLett.105.097001,Sasaki2015.PhysicaC.514.206}.
The ovals are Fermi surfaces, whose surface color represents the gap magnitude with black indicating $\Delta_{\bm{k}} = 0$.
\label{fig1}
}
\end{figure}

Fascinating features of superconductivity, such as zero resistivity or macroscopic quantum-mechanical coherence, are mostly governed by the superconducting (SC) gap $\Delta$, or equivalently, by the SC wavefunction. 
Study of ``unconventional'' superconductivity, exhibiting interesting SC phenomena originating from complex form of $\Delta$ in the reciprocal space, has been one of the most exciting topics in condensed-matter physics.
Unconventional superconductivity can be defined and classified in several ways, but a most rigorous way is in terms of the spontaneous symmetry breaking of the SC wavefunction with respect to the crystalline and other symmetries of the system~\cite{Sigrist1991.RevModPhys.63.239}.
In particular, most of unconventional SC states are accompanied by the spontaneous rotational symmetry breaking (RSB).
When spontaneous RSB occurs only in the phase factor of the macroscopic SC wavefunction, phase-sensitive junction  techniques are required to detect it~\cite{Tsuei2000.RevModPhys.72.969,Nelson2004.Science.306.1151}.
In contrast, when spontaneous RSB occurs in the {\em amplitude} of the wavefunction as illustrated in Fig.~\ref{fig1}A, it emerges even in bulk thermodynamic quantities.
Such superconductivity can be termed ``nematic'' superconductivity~\cite{Fu2014.PhysRevB.90.100509}, in analogy to liquid-crystal phases with spontaneous RSB of molecular orientations.
Nematic natures of {\em normal-state} (non-SC) conduction electron have been recently found in several systems~\cite{Ando2002.PhysRevLett.88.137005,Borzi2007.Science.315.214,Kasahara2012.Nature.486.382,Okazaki2011.Science.331.439}, but not yet reported for SC states (see Supplementary Text \ref{sec:other_SCs} for details).

Among known unconventional superconductors, the copper-doped topological insulator \cbs~\cite{Hor2010.PhysRevLett.104.057001}, consisting of triangular-lattice layers of Bi and Se with intercalated Cu between layers (Figs.~\ref{fig1}B and C), is rather unique. 
In addition to the ordinary $s$-wave SC state $\Delta_1$, possible unconventional odd-parity SC states, labeled as $\Delta_2$, $\Delta_3$, $\Delta_{4x}$, and $\Delta_{4y}$, originating from strong spin-orbit interactions and multi-orbital nature have been proposed (Fig.~\ref{fig1}D)~\cite{Fu2010.PhysRevLett.105.097001,Ando2015.AnnuRevCondensMatterPhys.6.361,Sasaki2015.PhysicaC.514.206}.
These odd-parity states can be also categorised as topological SC states, which are accompanied by stable surface states originating from the non-trivial topology of the SC wavefunction.
Among these states, the $\Delta_{4x}$ and $\Delta_{4y}$ states are predicted to be nematic SC states with a non-zero nematic order parameter~\cite{Fu2014.PhysRevB.90.100509}, accompanied by two-fold in-plane anisotropy in the SC gap amplitude and in bulk properties~\cite{Nagai2012.PhysRevB.86.094507}, breaking the six-fold symmetry expected for the lattice.
Experimentally, the zero-bias conductance peak observed in the point-contact spectroscopy, indicating existence of unusual surface states, evidences topological superconductivity~\cite{Sasaki2011.PhysRevLett.107.217001}. 
On the other hand, the scanning-tunneling microscopy (STM) experiment on the $ab$ plane revealed $s$-wave-like tunneling spectra~\cite{Levy2013.PhysRevLett.110.117001}, which are later found to be actually inconsistent with the $s$-wave ($\Delta_1$) scenario~\cite{Mizushima2014.PhysRevB.90.184516}.
Rather, it is proposed that the STM spectra may be explained within the topological superconductivity scenario by taking into account the possible quasi-two-dimensional (Q2D) nature of the Fermi surface~\cite{Lahoud2013.PhysRevB.88.195107,Ando2015.AnnuRevCondensMatterPhys.6.361}.
More recently,  by the nuclear-magnetic resonance (NMR), spin-rotational symmetry is revealed to be broken in the SC state~\cite{Zheng2015.unpublished}, suggesting realization of the $\Delta_{4x}$ or $\Delta_{4y}$ states.

In this Report, based on the field-angle-resolved high-resolution specific-heat $\C$ measurements of single-crystalline \cbs\ ($\Tc \approx 3.2$~K)~\cite{MaterialMethods}, we report thermodynamic evidence for the spontaneous RSB in the SC gap amplitude for the first time among any known superconductors. 
We revealed clear two-fold oscillation in the in-plane field-angle dependence of $C/T$ and in the upper critical field $\Hcc$, breaking the six-fold rotational symmetry expected for the crystal lattice.
We further obtained evidence for the $\Delta_{4y}$ state, with gap minima (or nodes) along one of the Bi-Bi bonding directions.
Our results unambiguously show that \cbs\ belongs to a new class of materials with odd-parity nematicity and topological superconductivity.
We emphasize that the conclusion holds irrespectively of the dimensionality of the actual normal (N) state electronic structure (see Supplementary Text \ref{sec:Delta4_discussion}).

\begin{figure}
\begin{center}
\includegraphics[width=16cm]{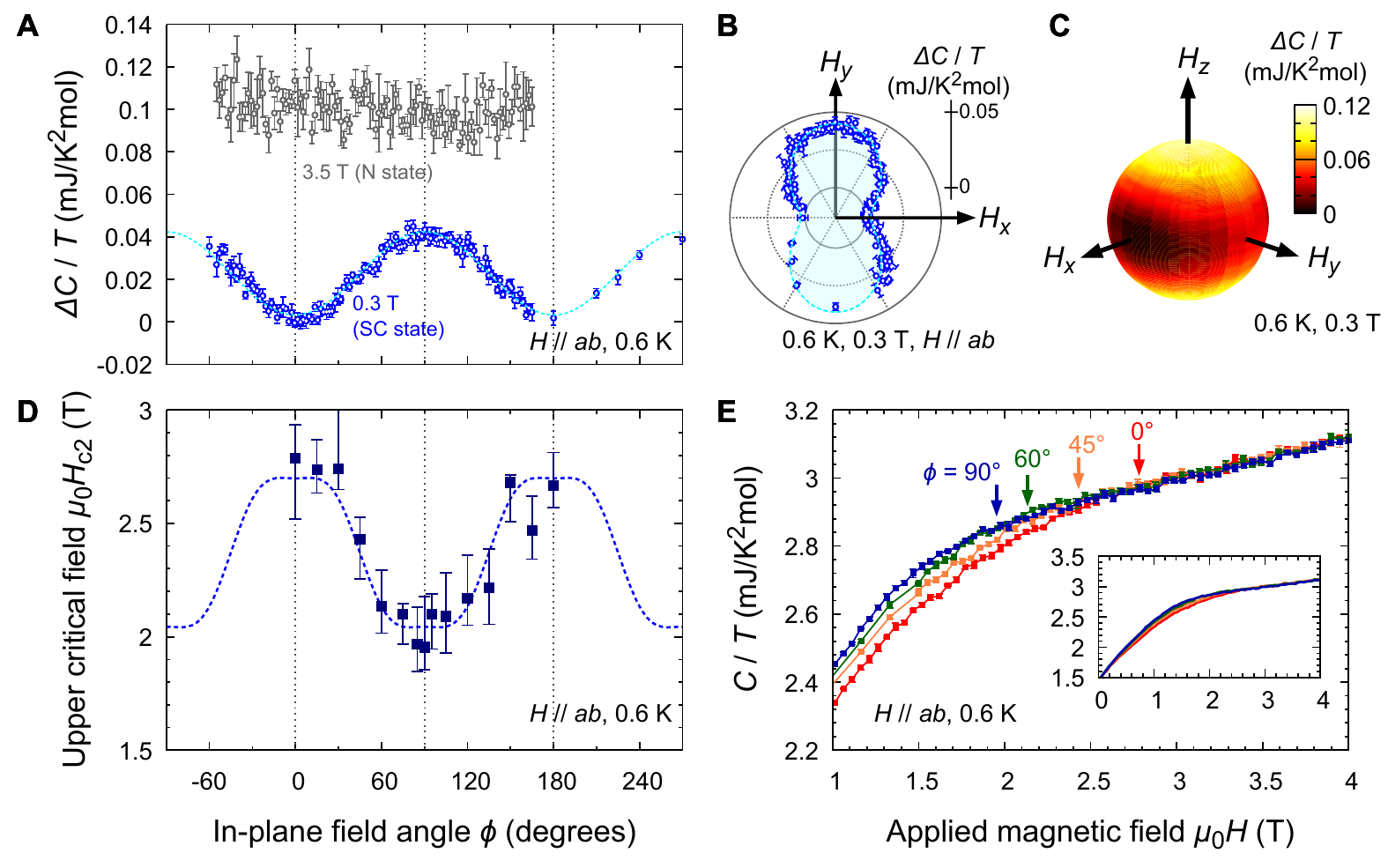}
\end{center}
\caption{ 
\baselineskip24pt
{\bf Evidence for nematic superconductivity in \cbsbm.}
{\bf A}. In-plane magnetic-field-angular oscillation of the specific heat at 0.6~K in the SC state (0.3~T, blue points) compared to the data in the N state (3.5~T, gray points). Here, $\Delta C(\phi)/T$ is defined as $C(\phi)/T - C(\bm{H}\parallel \bm{x})/T$ and the 3.5-T data is vertically shifted by 0.1~mJ/K\sps{2}mol. 
The broken curve is the fitting result with $\Delta C(\phi)/T = A_0 + A_2 \cos(2\phi)$. 
{\bf B}. Polar plot of $\Delta C(\phi)/T$ at 0.6~K and 0.3~T ($\bm{H}\parallel ab$), together with the fitting result. 
The polar angle of a data point in this 2D plot corresponds to the azimuthal field angle $\phi$ and the distance from the origin indicates the magnitude of $\Delta C/T$.
{\bf C}. Color map of $\Delta C(\theta,\phi)/T$ on a sphere in the $H_x$-$H_y$-$H_z$ space, based on the $\theta$-sweep data at 0.6~K and 0.3~T (see also Fig.~\ref{fig:Theta_phi_compare}A).
{\bf D}. In-plane anisotropy of $\Hcc$ at 0.6~K. The broken curve presents the result of fitting with $\Hcc(\phi) = H_0 + H_2\cos(2\phi) + H_6\cos(6\phi)$.
{\bf E}. In-plane magnetic-field dependence of the specific heat for various $\phi$. The arrows mark $\Hcc$. The inset presents data in the whole field range. 
\label{fig2}
}
\end{figure}

In Fig.~\ref{fig2}A, we compare the in-plane field-angle $\phi$ dependence of $\C/T$ of Sample~\#1 in the SC and N states. 
Here, as shown in Fig.~\ref{fig1}C, we define the $x$ axis as one of the six equivalent Bi-Bi bond directions within the $ab$ plane, the $y$ axis as the direction perpendicular to $\bm{x}$ within the plane, and $\phi$ as the azimuthal angle of the field with respect to $\bm{x}$.
In addition, the $z$ axis is parallel to the $c$ axis and the angle $\theta$ is the polar angle with respect to $\bm{z}$.
Although $\C/T$ is independent of $\phi$ in the N state, $\C(\phi)/T$ in the SC state unexpectedly exhibits clear two-fold oscillation.
For the rhombohedral $R\bar{3}d$ crystal symmetry of \cbs, $\C(\phi)/T$ should exhibit six-fold oscillation. 
Thus, the observed two-fold oscillation in $\C(\phi)/T$ clearly breaks the rotational symmetry of the underlying lattice.
The RSB is more easily recognized in the polar plot of $\C(\phi)/T$ in Fig.~\ref{fig2}B.
A possible extrinsic origin for such RSB is the field misalignment with respect to the $ab$ plane.
To inspect this possibility, we measured the polar-angle $\theta$ dependence of $\C$ for various $\phi$, presented as a surface color plot in Fig.~\ref{fig2}C (also see Fig.~\ref{fig:Theta_phi_compare}A). 
Evidently, $\C(\theta)/T$ exhibit minima at $\theta=90\deg$ for any $\phi$, excluding possibility of field misalignment. 
In addition, the two-fold oscillation has been reproduced in several samples (Fig.~\ref{fig:different_samples}).
Furthermore, one sample (\#3) exhibits shifted and smaller oscillation indicating existence of multiple ``nematic domains'', which manifest the spontaneous nature of the RSB (see Supplementary Text \ref{sec:domains}).
Therefore, the rotational symmetry of the lattice is intrinsically and spontaneously broken in the SC state, evidencing nematic superconductivity in \cbs.

Next, we discuss the in-plane anisotropy of $\Hcc$ presented in Fig.~\ref{fig2}D. 
In Fig.~\ref{fig2}E, we present the field-strength dependence of $\C/T$ at 0.6~K for various in-plane field directions.
Clearly, $\C(H)/T$ curves again do not obey the expected six-fold rotational symmetry: the curves are substantially different between $\phi=0\deg$ and $60\deg$.
We here define $\Hcc$ as the onset of deviation from the linear field dependence in the N state (see Supplementary Text \ref{sec:Hc2}).
The obtained $\Hcc(\phi)$ (Fig.~\ref{fig2}D) is clearly dominated by two-fold oscillation.
Indeed, by fitting $\Hcc(\phi)$ with $H_0 + H_2\cos(2\phi) + H_6\cos(6\phi)$, we obtain $\mu_0H_0 = 2.37\pm 0.03$~T, $\mu_0 H_2 = 0.37\pm 0.04 $~T, and $\mu_0 H_6 = -0.05\pm 0.04$~T.
Interestingly, $H_2$ is as large as 16\% of $H_0$.
This striking in-plane $\Hcc$ anisotropy not only  support the nematic SC state of \cbs, but also indicate existence of a single nematic domain in this sample (see Supplementary Text \ref{sec:domains}).

Among the proposed SC states for \cbs, only the $\Delta_{4x}$ and $\Delta_{4y}$ states spontaneously break the in-plane rotational symmetry~\cite{Nagai2012.PhysRevB.86.094507,Fu2014.PhysRevB.90.100509}.
Thus, the observed bulk nematicity provides strong evidence for the $\Delta_{4x}$ or $\Delta_{4y}$ states.
Furthermore, since these states belong to odd-parity topological SC states~\cite{Fu2010.PhysRevLett.105.097001}, our finding thermodynamically evidences that \cbs\ is indeed a topological superconductor.

\begin{figure}
\begin{center}
\includegraphics[width=16cm]{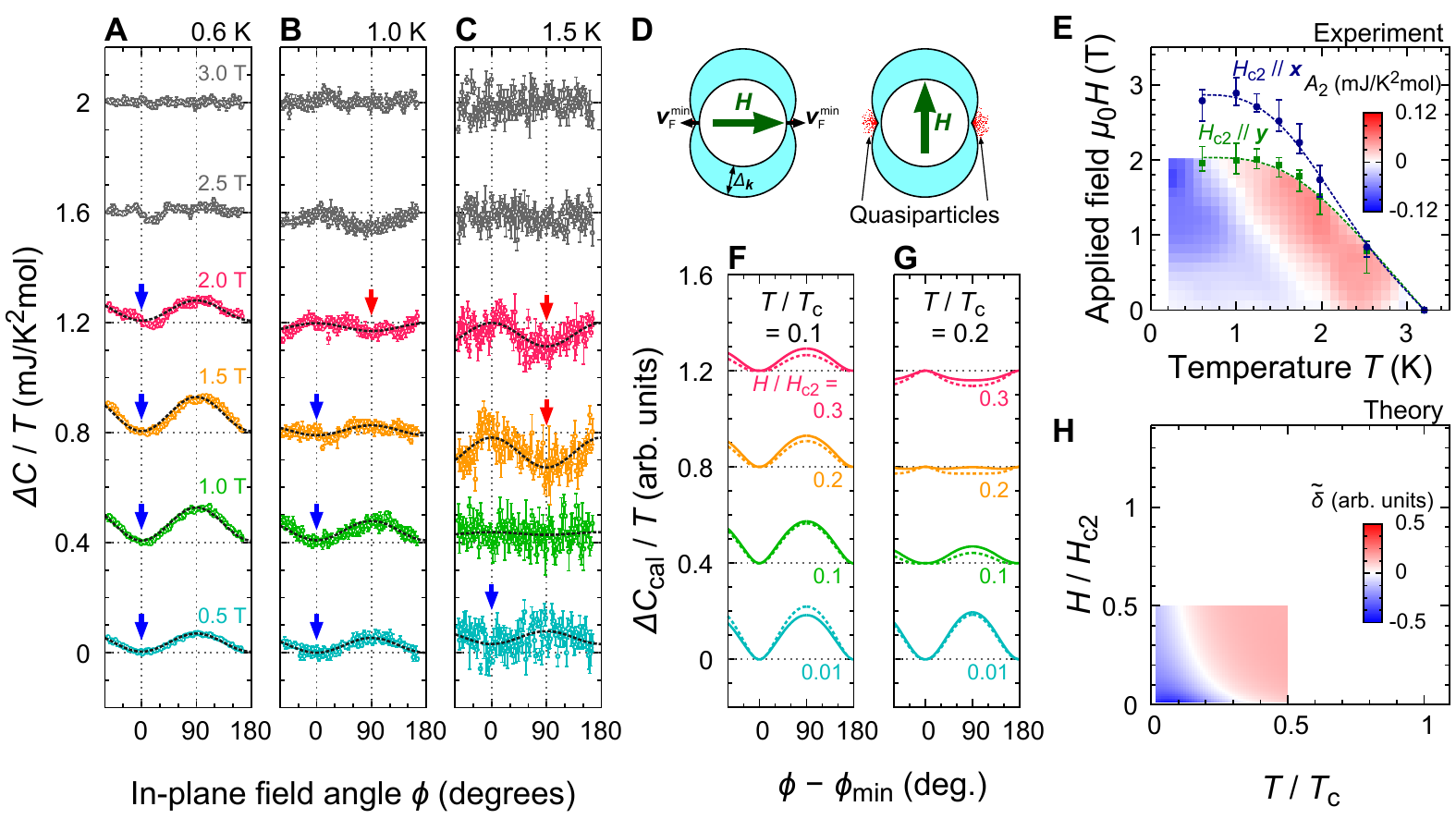}
\end{center}
\caption{
\baselineskip24pt
{\bf Superconducting gap structure of \cbsbm.}
{\bf A--C}. Experimental $\Delta\C(\phi)/T$ data compared to fitting results with $A_0 + A_2 \cos(2\phi)$ (broken curves). 
Each curve is shifted vertically by 0.4~mJ/K\sps{2}mol.
Notice that the sample is entirely in the N state at 3.0~T but is in the SC state near $\phi=0\deg$ at 2.5~T due to the $\Hcc$ anisotropy (see Fig.~\ref{fig2}E).
The vertical arrows indicate minimum positions in $\Delta\C(\phi)/T$.
{\bf D}. Schematics of the field-angle dependent quasiparticle excitation (red dots) in the low-temperature and low-field limit.
{\bf E}. Color map of $A_2$ within the SC phase. 
In the blue and red regions, $\Delta\C(\phi)/T$ exhibits a minimum at $\phi = 0\deg$ and $90\deg$, respectively.
The circles and squares indicate $\Hcc$ along $\bm{x}$ and $\bm{y}$.
{\bf F--G}. Calculated $\Delta\C\subm{cal}(\phi)/T\equiv C\subm{cal}(\phi)/T - C\subm{cal}(\phi\subm{min})/T$ with offsets.
Here, $C\subm{cal}$ is the calculated specific heat and $\phi\subm{min}$ is the field direction where $\bm{H} \parallel \vmin$.
The solid and broken curves are for point minima ($\Delta\subm{min}/\Delta\subm{max} = 0.1$) and for point nodes, respectively.
{\bf H}. Color map of $\tilde{\delta}\equiv \C\subm{cal}(\phi\subm{min})/\C\subm{cal}(\phi\subm{min}+90\deg) - 1$ for point-node gap.
In the blue and red regions, $\Delta\C\subm{cal}(\phi)/T$ exhibits a minimum and maximum at $\phi = \phi\subm{min}$, respectively.
\label{fig3}
}
\end{figure}

The two possible nematic states, the nodal $\Delta_{4x}$ state (with nodes along the $\bm{k}_y$ direction, protected by the mirror symmetry~\cite{Fu2014.PhysRevB.90.100509}) and the fully-gapped $\Delta_{4y}$ state (with gap minima along the $\bm{k}_x$ direction; also see Supplementary Text \ref{sec:Delta4_discussion}), can be distinguished by the position of the gap minima or nodes.
To this goal, we investigate $\C(\phi)/T$ of Sample~\#1, with a single nematic domain, in more detail.
When the SC gap has minima (including nodes), $\C/T$ exhibits oscillatory behavior as a function of field angle, because of the field-angle dependent quasiparticle excitations originating from the gap anisotropy~\cite{Vekhter1999,Sakakibara2007.JPhysSocJpn.76.051004.review}.
At low-temperature and low-field conditions, $\C/T$ exhibits minima when the field is parallel to the Fermi velocity at a gap minimum $\vmin$ as shown in Fig.~\ref{fig3}D.
Furthermore, it has been predicted and observed that, in addition to the $\C/T$ oscillation originating from $\Hcc$ anisotropy, the $\C/T$ oscillations exhibit sign changes depending on temperature and field conditions: for example, at intermediate temperatures, $\C/T$ exhibits {\em maxima} for $\bm{H}\parallel \vmin$~\cite{VorontsovA2006.PhysRevLett.96.237001,An2010.PhysRevLett.104.037002}.
Thus, detailed experiments as well as comparison with theoretical calculations are required to conclude the gap structure.

Figures~\ref{fig3}A-C represent the observed $\C(\phi)/T$ curves in various conditions.
The two-fold oscillation with minimum at $\phi=0\deg$ ($\bm{H}\parallel \bm{x}$) is observed at 0.6~K in the SC state.
However, at higher temperatures, the oscillation inverts sign above $\sim 1.5$--2.0~T at 1.0~K and $\sim 1.0$~T at 1.5~K, exhibiting {\em maximum} at $\phi=0\deg$.
The temperature and field dependence of the oscillation prefactor $A_2$ is summarized in Fig.~\ref{fig3}E as a color plot.
The boundary between positive and negative $A_2$ exists within the SC phase.
These observations are compared with theoretical calculations based on the Kramer-Pesch approximation~\cite{MaterialMethods} in Figs.~\ref{fig3}F and G.
Here, we assume a gap structure with gap minima or point nodes along an in-plane direction $\phi=\phi\subm{min}$ on a spherical Fermi surface.
The calculated $\C(\phi)/T$ curves for both cases of gap minima and nodes resemble the observed ones quite well: they exhibit two-fold oscillation with minimum for $\phi=\phi\subm{min}$ at low temperatures and low fields,
and reversed oscillation with maximum for $\phi=\phi\subm{min}$ at elevated temperatures. 
As shown in Fig.~\ref{fig3}H, the phase-inversion line passes at $T/\Tc \sim 0.35$ near $H/\Hcc=0$ and at $T/\Tc \sim 0.15$ at $H/\Hcc = 0.25$,
again qualitatively similar to the observation (Fig.~\ref{fig3}E).

From these agreements between experiment and theory, we conclude that the SC gap of \cbs\ is $\Delta_{4y}$, possessing gap minima or nodes lying along the $\bm{k}_x$ direction.
Although it is not straightforward to distinguish gap minima or nodes only from our data, it is more natural to expect that the $\Delta_{4y}$ state is fully gapped to have gap minima, due to symmetry and energetic reasons~\cite{Fu2014.PhysRevB.90.100509} (see Supplementary Text \ref{sec:Delta4_discussion}).

To summarize, we observed spontaneous RSB in the specific heat and upper critical field of \cbs.
These thermodynamic results places \cbs\ to a novel class of materials simultaneously exhibiting nematic and topological superconductivity.
The odd-parity nematic SC state in \cbs, accompanied by nematicity in the macroscopically coherent odd-parity quantum-mechanical wavefunction, is clearly distinct from known nematic states in liquid crystals and non-SC electrons. 
It is interesting to investigate unusual consequences of the odd-parity nematic SC ordering such as topological defects and collective modes.

\noindent
{\bf Acknowledgments}\\
We acknowledge T. ~Watashige, S.~Kasahara, and Y.~Kasahara for their technical assistance; L.~Fu, M.~Ueda, Y.~Yanase, J.~Yamamoto, Y.~Matsuda, A.~Yamakage, Y.~Tanaka, and T.~Mizushima, for fruitful discussion.
This work was supported by JSPS Grant-in-Aids for
Scientific Research on Innovative Areas on ``Topological Quantum Phenomena'' (KAKENHI 22103002, 22103004) and ``Topological Materials Science''  (KAKENHI 15H05852, 15H05853), and by JSPS Grant-in-Aids KAKENHI 26287078 and 26800197.
This study was designed by S.Y., Y.A., and Y.M..
K.T. and S.Y. performed specific-heat measurements and analyses, with an assistance of S.N. and a guidance of Y.M.. 
Z.W., K.S., and Y.A. grew single crystalline samples and characterized them.
Y.N. performed theoretical calculation.
The manuscript was prepared mainly by S.Y. and K.T., based on discussion among all authors.
All authors declare there is no competing interests regarding this work.

\vspace{1em}
\noindent
{\bf Supplementary Materials}\\
    www.sciencemag.org\\
    Materials and Methods\\
    Supplementary Text\\
    Figs. S1 to S10\\
    Table S1\\
    References (24-39)

\clearpage

%

%

%


\clearpage
\appendix
\renewcommand{\theequation}{S\arabic{equation}}
\setcounter{equation}{0}
\renewcommand{\thefigure}{S\arabic{figure}}
\setcounter{figure}{0}
\renewcommand{\thetable}{S\arabic{table}}
\setcounter{table}{0}
\renewcommand{\thepage}{S\arabic{page}}
\setcounter{page}{1}
\renewcommand{\thesubsection}{S\arabic{subsection}}
\setcounter{equation}{0}

\begin{center}

\vspace{0.3cm}

{\Large Supplementary Materials for\\[0.3cm]
\bfseries{Thermodynamic evidence\\for nematic superconductivity in Cu$_{\bm{x}}$Bi$_{\bm{2}}$Se$_{\bm{3}}$}}

\vspace{0.5cm}

{\large
Shingo~Yonezawa$^{1\ast}$, Kengo~Tajiri$^{1}$, Suguru~Nakata$^{1}$, Yuki~Nagai$^{2}$, \\ Zhiwei~Wang$^{3,4}$, Kouji~Segawa$^{3,5}$,
Yoichi~Ando$^{3,4}$, Yoshiteru~Maeno$^1$\\
}

\vspace{0.4cm}

\normalsize{$^{1}$Department of Physics, Graduate School of Science, Kyoto University,}\\
\normalsize{Kitashirakawa-Oiwake-Cho, Sakyo-Ku, Kyoto 606-8502, Japan}\\[0.2cm]
\normalsize{$^{2}$CCSE, Japan Atomic Energy Agency,}\\ \normalsize{178-4-4, Wakashiba, Kashiwa, Chiba 277-0871, Japan}\\[0.2cm]
\normalsize{$^{3}$The Institute of Scientific and Industrial Research, Osaka University,}\\
\normalsize{Mihogaoka 8-1, Ibaraki, Osaka 567-0047, Japan}\\[0.2cm]
\normalsize{$^{4}$Institute of Physics II, University of Cologne,}\\
\normalsize{Z\"{u}lpicher Str. 77, 50937 K\"{o}ln, Germany}\\[0.2cm]
\normalsize{$^{5}$Department of Physics, Graduate School of Science, Kyoto Sangyo University,}\\
\normalsize{Motoyama, Kamigamo, Kita-ku, Kyoto 603-8555 Japan}\\[0.3cm]
\normalsize{$^\ast$To whom correspondence should be addressed; }\\
\normalsize{E-mail: yonezawa@scphys.kyoto-u.ac.jp}

\end{center}

\vspace{0.5cm}

\noindent
{\bfseries{This PDF file includes:}}\\
Materials and Methods\\
Supplementary Text\\
Figs.~S1 to S10\\
Table~S1\\

\clearpage

\baselineskip 16pt 

\section*{Materials and Methods}

\begin{figure}[tb]
\begin{center}
\includegraphics[width=16cm]{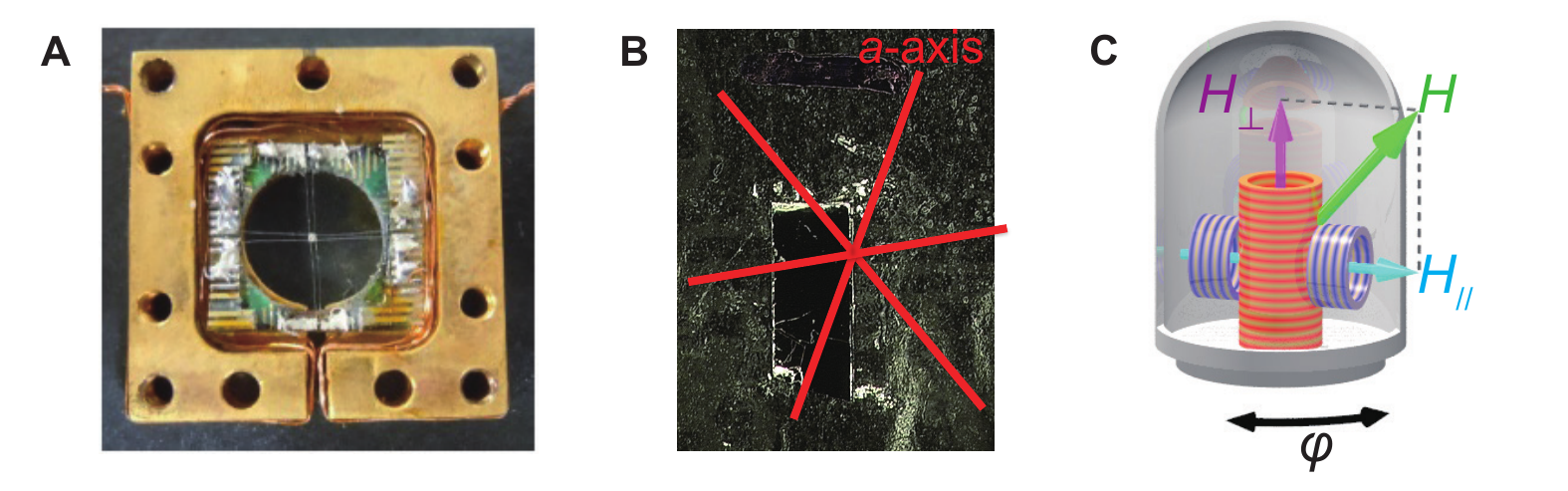}
\end{center}
\caption{
{\bf Equipments and samples used in this study.}
{\bf A}. Hand-made high-resolution calorimeter used in this study.
{\bf B}. Photo of a single crystal (Sample~\#1). The crystalline $a$ axis is shown with the red lines.
{\bf C}. Schematic image of the vector magnet system~\cite{Deguchi2004RSI}. 
\label{fig:Method}
}
\end{figure}

Single crystals of \bs\ is grown by a conventional melt-growth method.
Single crystal samples of \CBS{x} were then obtained by intercalating Cu to single crystals of \BS\ by an electrochemical technique~\cite{Kriener2011.PhysRevLett.106.127004,Kriener2011.PhysRevB.84.054513}.
The value of the Cu content $x$ is determined by the total charge flow during the intercalation process, as well as by the mass change between before and after the intercalation.
The crystal axis directions were determined by Laue photos before the intercalation.
The onset $\Tc$ is checked by using the superconducting quantum interference device (SQUID) magnetometer (MPMS, Quantum Design).
After this process, the samples are stored in vacuum until they are mounted to the calorimeter.
In the present study, we used three samples, which are labeled Samples~\#1 (Fig.~\ref{fig:Method}B), \#2 and \#3. 
All samples are single crystals of rectangular shapes with $x\simeq 0.3$.
The characteristics of each sample are listed in Table~\ref{tab:samples}

\begin{table}[b]
\caption{{\bf Characteristics of the \cbs\ samples used in this study.}
\label{tab:samples}}
\begin{center}
{\small
\begin{tabular}{ccccccc}\hline
 & Weight& Size in the $ab$ plane& Thickness & Direction of the $a$ axis & $T_{\mathrm{c},M}$~${}^\dagger$ & $T_{\mathrm{c},C}$~${}^\ddagger$ \\ \hline\hline
\#1 & 13.3~mg &  4.0~$\times$~1.7~mm$^2$ & 0.25~mm & $20\pm 1\deg$ off the long edge & 3.6~K & 3.2~K\\
\#2 &  5.74~mg & 2.3~$\times$~1.3~mm$^2$  & 0.26~mm & parallel to the short edge  & 3.5~K & 1.3~K~${}^\ast$ \\
\#3 &  6.64~mg & 3.0~$\times$~1.2~mm$^2$  & 0.24~mm & parallel to the long edge  & 3.6~K & 3.5~K \\ \hline
\end{tabular}
}
\end{center}
{\footnotesize
\hspace{1cm}$^\dagger$Determined as the onset of the decrease in $M(T)$.

\hspace{1cm}$^\ddagger$Determined as the onset of the increase in $\C\subm{el}(T)/T$.

\hspace{1cm}$^\ast$This sample degraded during storage, possibly due to a leak in the container.
}
\end{table}

We used a ${}^3$He-${}^4$He dilution refrigerator (Kelvinox 25, Oxford Instruments) to cool down the samples.
We performed specific-heat measurement in the temperature range 0.09~K ${}\leq T \leq{}$ 4~K. 
We constructed a hand-made high-resolution calorimeter~\cite{Yonezawa2013.PhysRevLett.110.077003} shown in Fig.~\ref{fig:Method}A. 
In our calorimeter, a sample is sandwiched by a thermometer and a heater, both of which are made with RuO$_2$ chip resisters.
We mounted a sample inside a glove box with Ar atmosphere.
We measured the specific heat by using the AC method~\cite{Sullivan1968}: 
We applied AC current to the heater using a current source (6221, Keithley Instruments Inc.), and measured the resultant temperature modulation amplitude $T\subm{ac}$ and the phase shift $\phi\subm{ac}$ using lock-in amplifiers (SR830, Stanford Research Systems). 
The offset sample temperature $T$ is also recorded by another lock-in amplifier. 
Excitation current to the thermometer is applied using another current source.
For most of the data, the raw heat capacity $C\subm{raw}$ is then obtained as $C\subm{raw}  = [P/(2\omega_\mathrm{H} T\subm{ac})]\sin \phi\subm{ac}$, where $P$ is the AC heat flow produced by the heater and $\omega_\mathrm{H}$ is the frequency of the heater current.
Multiplication by the factor $\sin \phi\subm{ac}$ allows for more accurate evaluation of the heat capacity even for smaller frequencies~\cite{Velichkov1992.Cryogenics.32.285}.
Notice that the frequency of the temperature oscillation is twice higher than $\omega\subm{H}$.
We typically used $\omega\subm{H} = 1.3$~Hz and $T\subm{ac}/T \sim 1.5$--2.0\% for measurements.
For the temperature dependence of $\C/T$ below 0.6~K and at zero field (Fig.~\ref{fig:T-sweep}), we evaluated $C\subm{raw}$ by another method to improve the accuracy:
we measured the $\omega\subm{H}$ dependence of $T\subm{ac}$  and fitted $T\subm{ac}(\omega\subm{H})$ with the function $T\subm{ac}(\omega\subm{H})= [P/(4\omega\subm{H} C\subm{raw})][1 + (2\omega\subm{H} \tau_1)^2 + (2\omega\subm{H} \tau_2)^{-2}]^{-1/2}$ to obtain $C\subm{raw}$, where $\tau_1$ and $\tau_2$ are the external and internal relaxation times, respectively.
The heat capacity of the sample stage (addenda) was measured separately by using a piece of pure silver as a reference sample, and subtracted from the total heat capacity to extract the sample contribution.
The addenda contribution is typically 10-20\% of the total heat capacity.
We confirmed that the addenda does not exhibit detectable change in both $\theta$ and $\phi$ dependences.

We applied the magnetic field with a vector-magnet system~\cite{Deguchi2004RSI}, which consists of two orthogonal superconducting solenoids and a rotation stage as schematically shown in Fig.~\ref{fig:Method}C.
This system allows for a precise three-dimensional control of the field direction.
The magnetic field is aligned to the crystalline axes by making use of the anisotropy in $\Hcc$.
The precision and accuracy of the field alignment are approximately $1\deg$. 
Notice that this is worse than those achieved for more anisotropic superconductors such as Sr\sub{2}RuO\sub{4}~\cite{Yonezawa2013.PhysRevLett.110.077003}, because of the relatively small $\Hcc$ anisotropy of \cbs. 
Nevertheless, this small anisotropy of \cbs\ in turn makes the misalignment effect rather small, as explained later.

The field-angle-dependent heat capacity is calculated on the basis of the Kramer-Pesch approximation, which is appropriate in a low magnetic-field region~\cite{Nagai2011.PhysRevB.83.104523}. 
With the use of the quasiclassical framework, a Dirac Bogoliubov-de Gennes (BdG) Hamiltonian which describes a topological superconductivity with point-nodes derived from the first-principle calculations is reduced to a BdG Hamiltonian for spin-triplet $p$-wave superconductivity. 
The corresponding $\bm{d}$-vector is $\bm{d}(\kF) = (v_{\mathrm{F}z} \sin\phi\subm{N}, -v_{\mathrm{F}z}\cos\phi\subm{N}, v_{\mathrm{F}y} \cos\phi\subm{N}-v_{\mathrm{F}x}\sin\phi\subm{N})$. 
For this state, point nodes are located in the $\phi\subm{N}$ direction on the $ab$ plane~\cite{Nagai2014.JPhysSocJpn.83.063705}.
Here, we consider a three dimensional spherical Fermi surface. 
In the case of a fully-gapped order parameter with gap minima, we consider the gap $|\Delta(\bm{k})| = |\bm{d}(\kF)|(1-r) + |\bm{d}|\subm{max}r$, where $|\bm{d}|\subm{max}$ is the maximal value of $|\bm{d}(\kF)|$ and $r$ ($0<r<1$) corresponds to the ratio between the minimal and  maximal values of $|\Delta(\kF)|$.
We also performed calculation for a Q2D Fermi surface. We used a Q2D tight-binding model for the normal-state electronic band~\cite{Lahoud2013.PhysRevB.88.195107} and assumed a point-nodal gap~\cite{Hashimoto2014.SupercondSciTechnol.27.104002}, as schematically shown in the inset of Fig.~\ref{fig:cone}C.

\clearpage

\section*{Supplementary Text}

\subsection{Temperature dependence of the specific heat at zero field}
\label{sec:T-dep}

\begin{figure}[htb]
\begin{center}
\includegraphics[width=16cm]{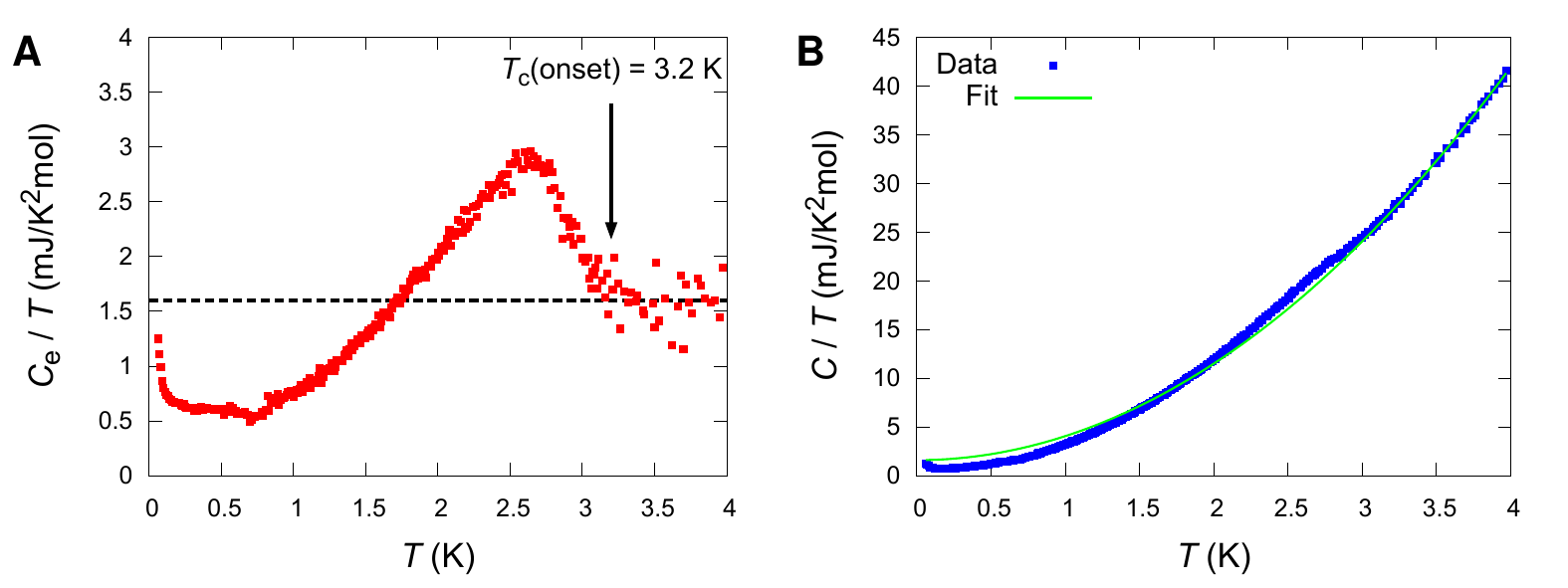}
\end{center}
\caption{
{\bf Temperature dependence of the specific heat of Sample~\#1.}
{\bf A}. Temperature dependence of the electronic specific heat $\C\subm{e}$ divided by temperature. The vertical arrow indicate the onset $\Tc$ ($\approx 3.2$~K) of the sample. 
{\bf B}. Specific heat of the sample including the phonon contribution. The green curve indicates the result of the fitting with the function $\C(T)/T = \gamma\subm{e} + \beta\subm{p}T^2$ ($\gamma\subm{e} =1.6$~mJ/K\sps{2}mol). 
\label{fig:T-sweep}
}
\end{figure}

In Fig.~\ref{fig:T-sweep}A, we present the electronic specific heat $\C\subm{e}$ divided by temperature as a function of temperature. 
The sample specific heat (shown in Fig.~\ref{fig:T-sweep}B) is obtained by subtracting the background contribution, and $\C\subm{e}$ is obtained by subtracting the phonon contribution obtained by the fitting of the data with $\C(T)/T = \gamma\subm{e} + \beta\subm{p}T^2$. 
For this fitting, $\gamma\subm{e}$ ($=1.6$~mJ/K\sps{2}mol) is determined so that the entropy balance between the SC and N states is satisfied, because there was large uncertainty in the value of $\gamma\subm{e}$ if we use ordinary fitting.
In $\C\subm{e}/T$, a clear bulk superconducting transition is observed with the onset $\Tc$ of 3.2~K.
The small upturn below 0.2~K is attributable to Schottky behavior of nuclear contributions (mainly by Bi with the nuclear spin $I = 9/2$).

\clearpage

\subsection{Comparison between $\bm{\theta}$ and $\bm{\phi}$ sweeps}
\label{sec:theta-sweep_compare}

\begin{figure}[htb]
\begin{center}
\includegraphics[width=16cm]{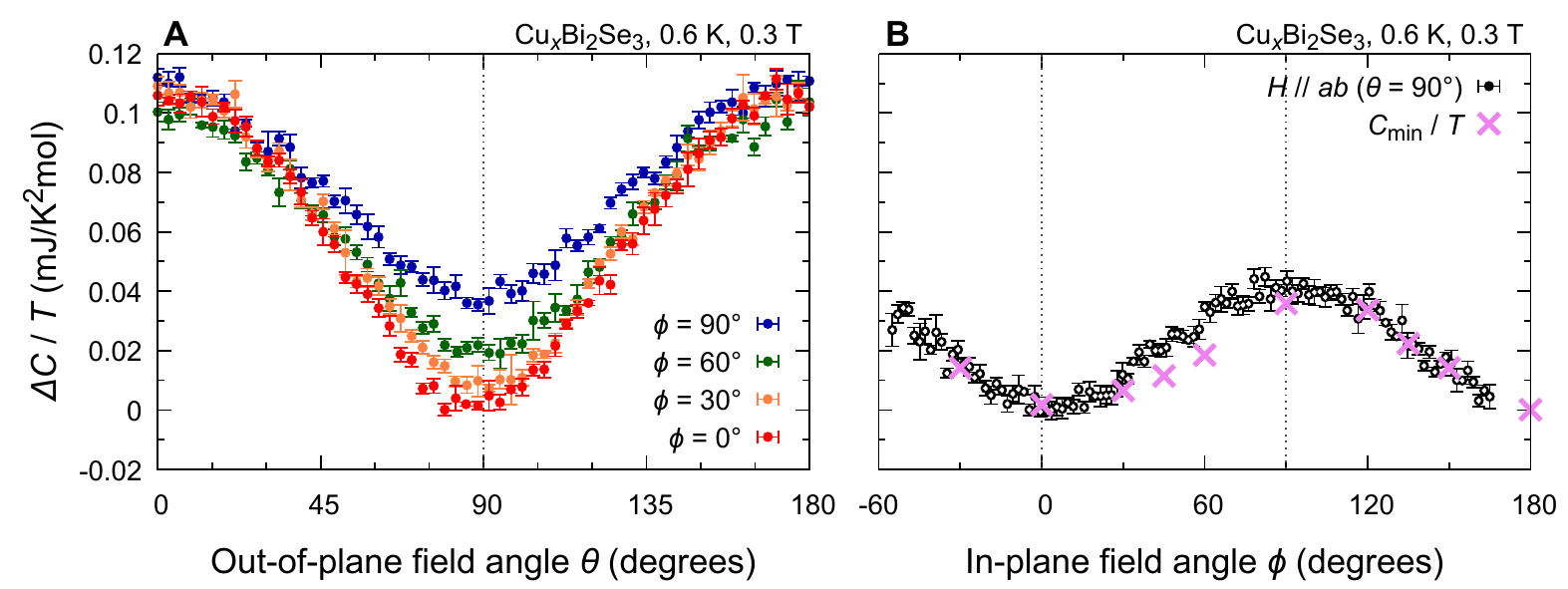}
\end{center}
\caption{
{\bf Comparison between the out-of-plane and in-plane field angle dependences of the specific heat.} 
{\bf A}. Out-of-plane field angle $\theta$ dependence of $\Delta\C(\phi, \theta)/T \equiv [\C(\phi, \theta) - \C(\phi=0\deg, \theta=90\deg)]/T$ at $\phi=0\deg$ (red circles), $30\deg$ (orange circles), $60\deg$ (green circles), and $90\deg$ (blue circles).
{\bf B}. In-plane field angle $\phi$ dependence of $\Delta\C/T$ at $\theta = 90\deg$ ($H\parallel ab$), 0.6~K, and 0.3~T (black circles). The purple crosses indicate $\C\subm{min}/T$, which are the minimal values in the $\theta$ sweeps represented in the panel A. The agreement between the two data sets indicate the absence of field misalignment effect in the present data.
The small discrepancy at $\phi=45\deg$ and $60\deg$ is mainly due to a small offset in $\Delta C(\theta)/T$ (see the panel A), rather than the field misalignment. 
\label{fig:Theta_phi_compare}
}
\end{figure}

To check the possibility of field misalignment, 
we compare $\theta$ dependence of $\C/T$ at various azimuthal angle $\phi$ in Fig.~\ref{fig:Theta_phi_compare}A. 
For all $\phi$, $\C(\theta)/T$ curves exhibit two-fold oscillation with minima at $\theta = 90\deg$, i.e. $\bm{H}\parallel ab$.
If there were field misalignment, $\theta\subm{min}$, where $\C(\theta)/T$ exhibits minimum, should deviate from $90\deg$, which is not the case.
Moreover, $\C\subm{min}/T \equiv \C(\theta\subm{min})/T$ values for various $\phi$ agrees with $\C(\phi)/T$ at $\theta = 90\deg$, as presented in Fig.~\ref{fig:Theta_phi_compare}B.
These facts give evidence that the field is correctly aligned to the crystalline $ab$ plane during the field-angle dependence measurements.
Thus, field misalignment cannot be the origin of the observed two-fold symmetry in $\C/T$ and $\Hcc$, and the spontaneous RSB is intrinsic to \cbs.

\clearpage

\subsection{Temperature and field dependence of the oscillation prefactor $\bm{A_2}$}
\label{sec:A2}

\begin{figure}[htb]
\begin{center}
\includegraphics[width=16cm]{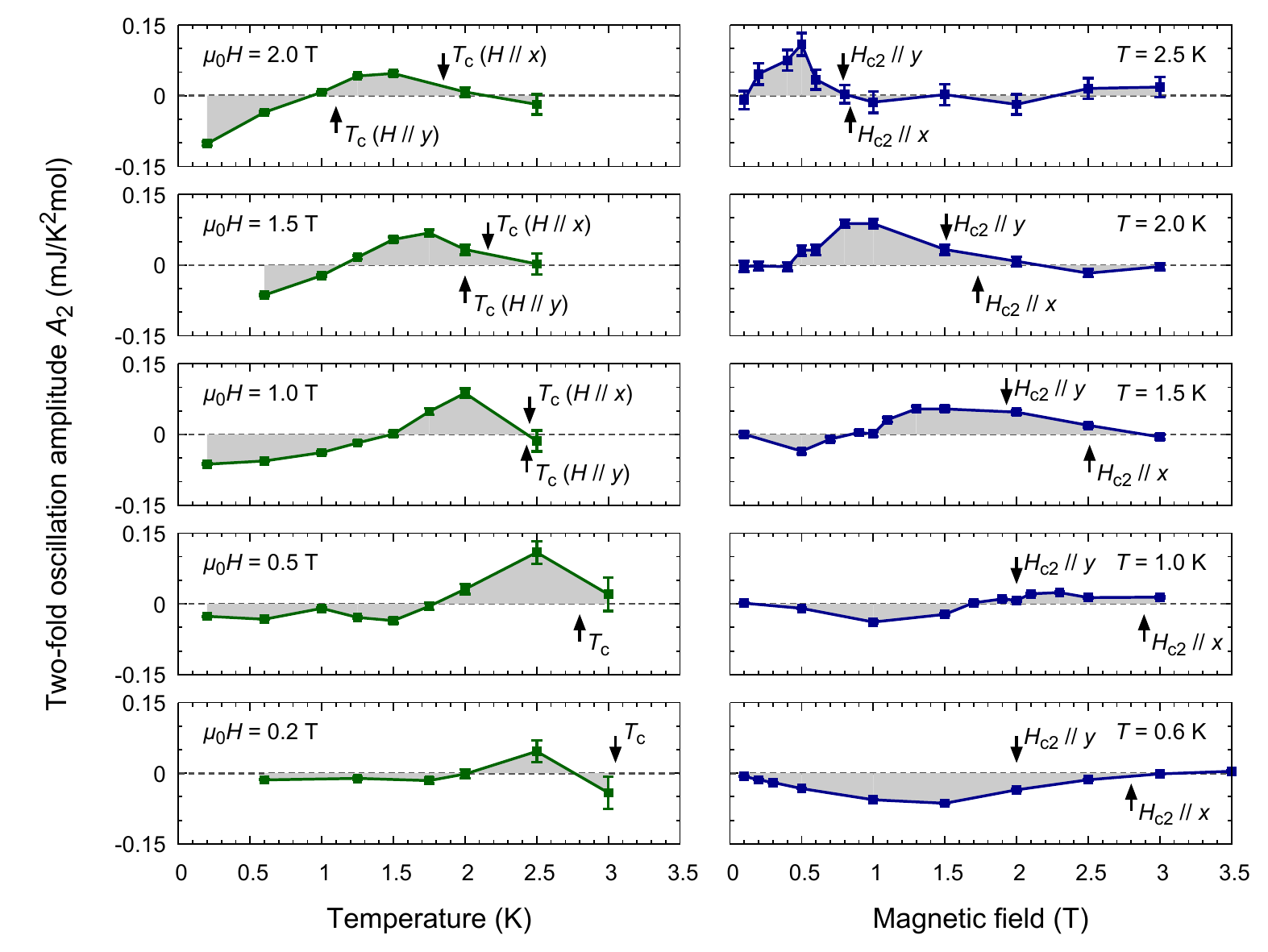}
\end{center}
\caption{
{\bf Temperature and field dependence of the two-fold oscillation prefactor.} 
In the left column, the two-fold oscillation prefactor $A_2$ at various fields is plotted as functions of temperature.
In the right column, $A_2$ at various temperatures is plotted as functions of field.
For each panel, $\Tc$ or $\Hcc$ are indicated with arrows.
Notice that, in the regions between $\Tc$ ($\bm{H}\parallel \bm{x}$) and $\Tc$ ($\bm{H}\parallel \bm{y}$) or in the regions between $\Hcc\parallel \bm{x}$ and $\Hcc\parallel \bm{y}$, the obtained $A_2$ is affected by partial destruction of superconductivity due to the in-plane $\Hcc$ anisotropy. 
\label{fig:A2_vs_T_and_H}
}
\end{figure}

In the left column of Fig.~\ref{fig:A2_vs_T_and_H}, we plot the prefactor $A_2$ of the two-fold $\C(\phi)/T$ oscillation as functions of temperature at various fields for Sample~\#1. 
These graphs correspond to horizontal cuts of the color plot in Fig.~\ref{fig3}E.
The inversion of the oscillation phase is clearly seen at around 1--2~K, depending on the field values.
Similarly, we plot $A_2$ as functions of magnetic field in the right column of Fig.~\ref{fig:A2_vs_T_and_H}.
The inversion is observed above 1.0~K but not observed at lower temperatures.
These data are consistent with the color plot.

\clearpage

\subsection{Specific-heat oscillation at the lowest temperature}
\label{sec:A2_low-T}

\begin{figure}[htb]
\begin{center}
\includegraphics[width=8cm]{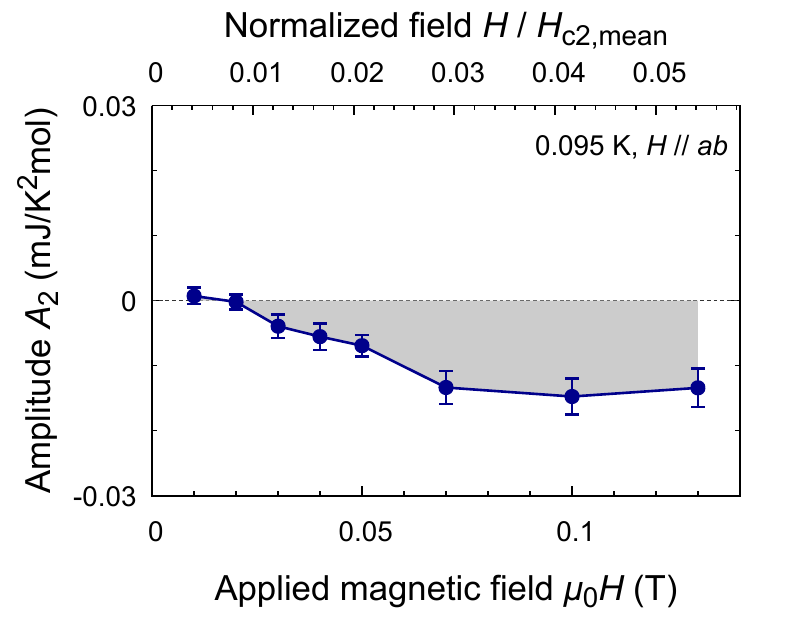}
\end{center}
\caption{
{\bf Two-fold oscillation prefactor at low-temperature and low-field conditions.}
The two-fold oscillation prefactor $A_2$ at 0.095~K is plotted against applied magnetic field. On the top horizontal axis, the value of the normalized field $H/H\subm{c2,mean}$ is shown. Here, $H\subm{c2,mean}$ is the average of $\Hcc$ for the $x$ and $y$ axes ($\mu_0H\subm{c2,mean} = 2.4$~T).
\label{fig:A2_vs_H_low-T}
}
\end{figure}

In Fig.~\ref{fig:A2_vs_H_low-T}, we plot $A_2$ measured at the lowest temperature (0.095~K, corresponding to $T/\Tc = 0.03$) as a function of field.
Clear oscillation with finite and negative $A_2$ (corresponding to oscillations with minima at $\bm{H}\parallel \bm{x}$, those observed in the blue region in the phase diagram in Fig.~\ref{fig3}E) was observed down to $\approx 0.03$~T, corresponding to $H/\Hcc\sim 0.015$.
These field and temperature conditions are only a few percents of $\Hcc$ and $\Tc$, respectively.
This fact supports our assignment that the observed specific-heat oscillation with minima for $\bm{H}\parallel \bm{x}$ is of the ``proper'' Volovik type (i.e. specific-heat minima when $H\parallel \vnode$).

\clearpage

\subsection{Specific heat under ``cone''-type field rotation}
\label{sec:cone}

\begin{figure}[htbp]
\begin{center}
\includegraphics[width=16cm]{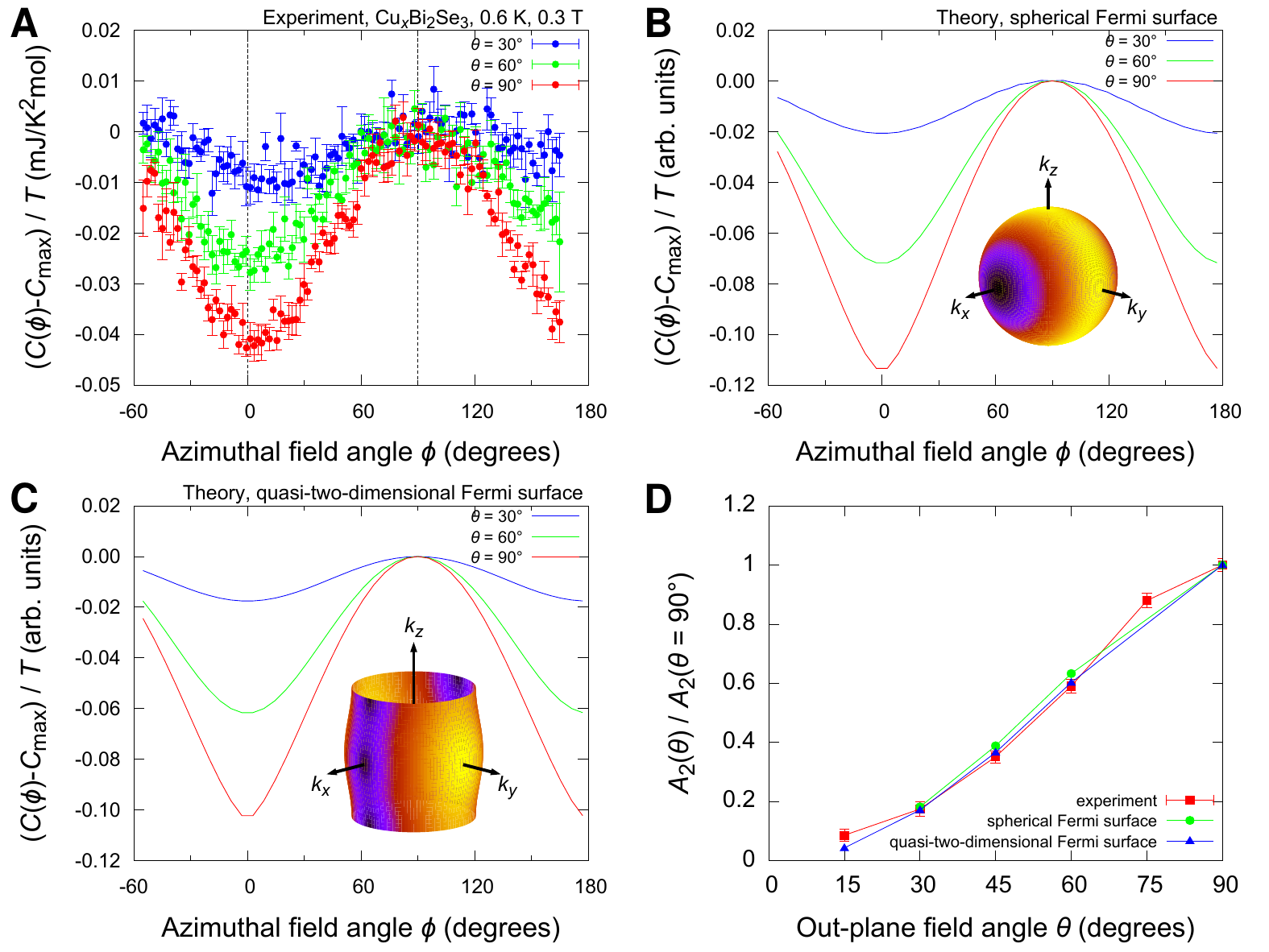}
\end{center}
\caption{
{\bf Specific-heat oscillation for cone-type field rotation.}
{\bf A}. Specific heat oscillation of \cbs\ as a function of the azimuthal field angle $\phi$ under the magnetic fields of the polar angles $\theta = 30\deg$, $60\deg$, and $90\deg$, measured at 0.6~K and 0.3~T. 
{\bf B}. Theoretical specific heat oscillation at various $\theta$, calculated for a point-nodal model on a spherical Fermi surface.
{\bf C}. Theoretical specific heat oscillation at various $\theta$, calculated for a point-nodal model on a Q2D Fermi surface~\cite{Lahoud2013.PhysRevB.88.195107,Hashimoto2014.SupercondSciTechnol.27.104002}.
In {\bf B} and {\bf C}, the Fermi surface and gap structure used for this calculation is shown in the inset: The color on the Fermi surface presents $k$-dependent gap amplitude with black corresponding to $\Delta = 0$.
{\bf D}. Dependence of the oscillation prefactor $A_2$ normalized by $A_2(\theta=90\deg)$ on the out-of-plane field angle $\theta$, compared with corresponding theoretical values.
\label{fig:cone}
}
\end{figure}

We comment here that the Fermi surface of superconducting \cbs\ may have a cylindrical shape, as revealed by angle-resolved photoemission and quantum oscillation studies~\cite{Lahoud2013.PhysRevB.88.195107}.
In such a quasi-two-dimensional (Q2D) Fermi surface, additional gap minima or nodes appears at a Brillouin-zone boundary along the $k_z$ direction~\cite{Hashimoto2014.SupercondSciTechnol.27.104002}, as shown in the inset of Fig.~\ref{fig:cone}C.
We thus performed calculation for the Q2D scenario and revealed that $\C/T$ oscillation for the Q2D case is quite similar to that for the 3D case.

In addition, hoping to distinguish the two possibilities, we measured $\C/T$ under ``cone''-type field rotation, i.e. $\phi$ dependence with $\theta \ne 90\deg$.
In Fig.~\ref{fig:cone}A, we present $\C(\phi)/T$ curves at 0.3~T and 0.6~K with various polar angle $\theta$.
The oscillation amplitude decreases as the field is tilted away from the $ab$ plane. 
The amplitude at $\theta = 45\deg$ is 1/3 of that of $\theta = 90\deg$ as plotted in Fig.~\ref{fig:cone}D.

These results should be compared with theoretical calculations.
In Figs.~\ref{fig:cone}B and C, $\C(\phi)/T$ curves calculated for various $\theta$ and for $H/\Hcc = 0.01$ and $T/\Tc =0.01$ are presented. 
The curves in the panel B is obtained for a point-nodal model with a spherical Fermi surface, whereas those in C are for a point-nodal model with a Q2D Fermi surface proposed in Ref.~\cite{Lahoud2013.PhysRevB.88.195107}.
The $\theta$ dependences of the oscillation for both models resemble that observed experimentally similarly well.
Indeed, the reduction of the oscillation amplitude with decreasing $\theta$ excellently agrees with the experiment, as shown in Fig.~\ref{fig:cone}D.
Thus, although it is supported again that the observed behavior originates from the anisotropic gap structure, it is difficult to conclude which of the two models, three-dimensional or quasi-two-dimensional, are realized in the actual samples.

Nevertheless, we here emphasize that the conclusion on the realization of the nematic $\Delta_{4y}$ state holds in both 3D and Q2D cases (see Supplementary Text \ref{sec:Delta4_discussion} for details).

\clearpage

\subsection{Reproducibility in different samples}
\label{sec:reproducibility}

\begin{figure}[htb]
\begin{center}
\includegraphics[width=16cm]{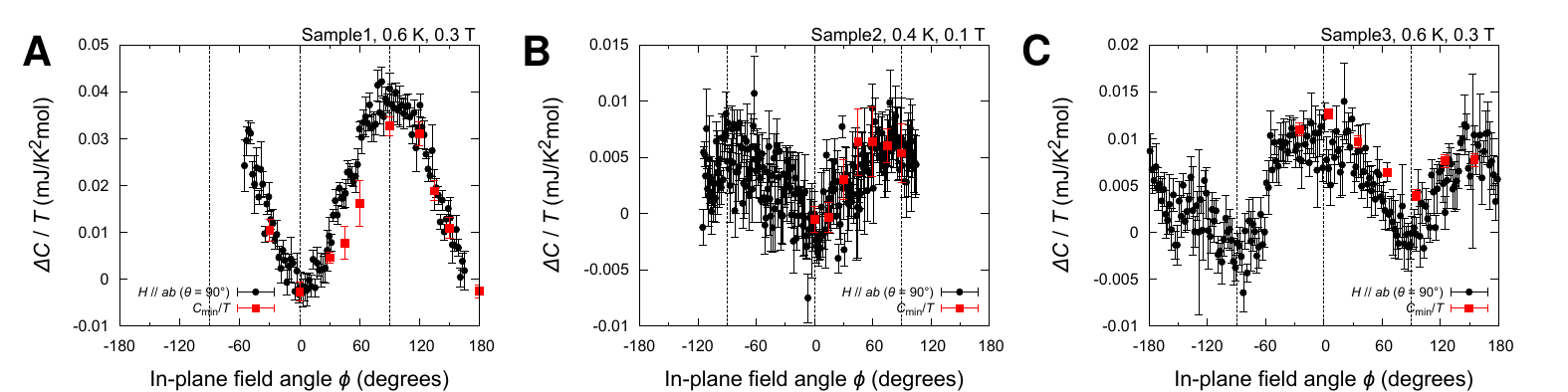}
\end{center}
\caption{
{\bf Reproducibility of the nematic behavior in various samples.}
{\bf A}. In-plane field angle $\phi$ dependence of $\Delta\C/T$ of Sample~\#1 at $\theta = 90\deg$ ($H\parallel ab$), 0.6~K, and 0.3~T (black circles), together with $\C\subm{min}/T$ of $\theta$-sweep curves (red squares). 
{\bf B}. Similar data of Sample~\#2 at 0.4~K and 0.1~T.
{\bf C}. Similar data of Sample~\#3 at 0.6~K and 0.3~T. Notice that the locations of the oscillation minima are $\phi=90\deg$, different from those of Sample~\#1. Also, the oscillation amplitude is substantially smaller. These discrepancy results presumably from existence of nematic domains in Sample~\#3.
\label{fig:different_samples}
}
\end{figure}

We checked the reproducibility of the observed nematic behavior.
In Fig.~\ref{fig:different_samples}, we compare data for Sample~\#2 and Sample~\#3 with those for Sample~\#1. 
For all samples, $\C(\phi)/T$ for $H\parallel ab$ exhibits two-fold oscillation.
Thus, the spontaneous RSB is ubiquitously observed in various samples.
This fact again supports that the nematic superconductivity is an intrinsic property of \cbs.

It should be noted that the minima of $\C(\phi)/T$ is located at $\phi=0\deg$ (i.e. $\bm{H}\parallel \bm{x}$) for Samples~\#1 and \#2 in the low-field and low-temperature region, whereas they are at $\phi=\pm 90\deg$ (i.e. $\bm{H}\parallel \bm{y}$) for Sample~\#3.
In addition, the oscillation amplitude of Sample~\#3 is only 1/3 of that of Sample~\#1, although they exhibit similar $\Tc$.
These apparent discrepancy is actually explained by assuming the existence of multiple ``nematic domains'' inside Sample~\#3, as explained in Supplementary Text \ref{sec:domains}.

\clearpage

\subsection{Possible nematic domains}
\label{sec:domains}

\begin{figure}[htbp]
\begin{center}
\includegraphics[width=16cm]{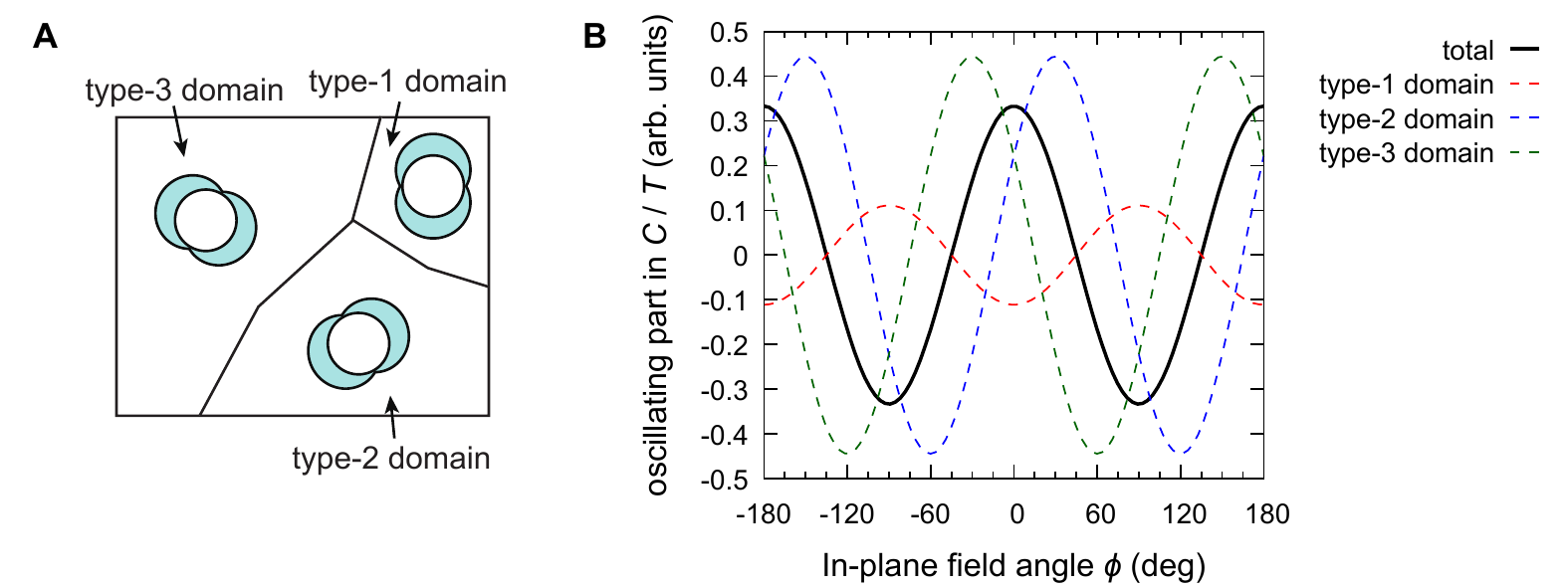}
\end{center}
\caption{
{\bf Model of multiple nematic domains.} 
{\bf A} Possible three types of domains for the $\Delta_{4y}$ state in \cbs.
{\bf B} Simulation of the specific-heat oscillation using Eq.~\eqref{eq:domain_model} in the case of $(v_1, v_2, v_3) = (1/9,\ 4/9,\ 4/9)$ and $A_2 < 0$.
The vertical scale is normalized by $|A_2|$. 
This simulation result well explains the behavior observed in Sample~\#3 (Fig.~\ref{fig:different_samples}C).
\label{fig:domains}
}
\end{figure}

We also comment here on the possible existence of ``nematic domains'' inside the sample.
Instead of having a single SC gap throughout the sample, the system may choose to form domains with three different gaps for the $\Delta_{4y}$ state in \cbs, as schematically depicted in Fig.~\ref{fig:domains}A.
We call these domains ``type-$i$'' ($i = 1,2,3$)  domains: the  type-1 domain with gap minima or nodal direction pointing along $\phi=0\deg$ and $180\deg$, the type-2 domain along $\phi=120\deg$ and $300\deg$, and the type-3 domain along $\phi=240\deg$ and $60\deg$. 

It is reasonable to assume that the type-$i$ domain should exhibit the $\C(\phi)/T$ oscillation as $v_i\left\{A_{2}\cos[2 (\phi-120\deg\times (i-1))] + A_0\right\}$, where $v_{i}$ is the relative volume fraction of each type of domain ($v_1 + v_2 + v_3 = 1$), $A_2$ is the ideal oscillation prefactor for a single-domain sample, and $A_0$ is a constant offset. 
The total $\C(\phi)/T$ should be expressed as 
\begin{equation}
\frac{\C(\phi)}{T} = A_2\sum_{i=1,2,3} v_{i}\cos[2 (\phi-120\deg\times (i-1))] + A_0.
\label{eq:domain_model}
\end{equation}
The oscillation amplitude and phase are both dependent on $v_i$. 
For example, if $v_1 = v_2 = v_3 = 1/3$, the oscillation completely vanishes.
If $v_1 = v_2 = 1/2$ and $v_3 = 0$, the oscillation should be expressed as
\begin{equation}
\frac{\C(\phi)}{T} = -\frac{A_2}{2} \cos[2(\phi-60\deg)] + A_0,
\end{equation}
exhibiting oscillation amplitude half of that of the single-domain case.
In addition, oscillation minima are now located at $\phi=-30\deg$ and $+150\deg$ in the ``simple Volovik'' regime ($A_2<0$), in clear contrast to the single-domain case where the minima should be located at integer multiples of $60\deg$.

As we already explained in \ref{sec:reproducibility}, the $\C(\phi)/T$ oscillation amplitude of Sample~\#3 is found to be approximately 1/3 of that for Sample~\#1, although $\Tc$ is nearly the same.
In addition, the minima of the two-fold $\C(\phi)/T$ oscillation are located at $\phi=\pm 90\deg$, rather than $\phi=0\deg$.
These behavior can be explained if we assume the existence of multiple nematic domains inside Sample~\#3 with the ratio $(v_1, v_2, v_3) = (1/9,\ 4/9,\ 4/9)$.
With this choice, as shown in Fig.~\ref{fig:domains}B, the total oscillation amplitude is indeed 1/3 of $|A_2|$ and the minima are located at $\pm 90\deg$ rather than $\phi=0\deg$ and $\phi=\pm 180\deg$.
Thus, it is quite likely that Sample~\#3 contains multiple nematic domains inside.
The domain formation provides strong evidence for spontaneous nature of the rotational symmetry breaking in \cbs.

It should be noted that Sample~\#1 consists of single domain, or is very close to single domain, because two-fold oscillation is observed also in $\Hcc$.
If multiple domains exist, $\Hcc$ anisotropy should be governed by the six-fold oscillation, since the measured $\Hcc$ detect the highest $\Hcc$ among different domains. 
Thus, our conclusion on the $\Delta_{4y}$ gap structure should be valid.

\subsection{Mechanism to fix the gap-minima direction}
\label{sec:mechanism_fix_gap}

Once the nematic superconducting order sets in, the system chooses a particular direction out of three equivalent $a$ axes as the directions of gap minima, resulting in spontaneous RSB.
This ``special'' $a$ axis is analogous to the ``director'' $\bm{\hat{n}}$ (headless vector) in the liquid-crystal nematic phases.
In an ideal case, the director $\bm{\hat{n}}$ should be chosen randomly.
Thus, the field direction for which $\Delta\C(\phi)$ exhibits minima should be altered when the sample is heated above $\Tc$ and cooled down again in zero field. 
However, in our experiments, such  ``random'' behavior has not been observed.
Therefore, there should be a mechanism to fix the director.
One possibility is that the director $\bm{\hat{n}}$ is determined by sample edges.
Another possibility is that a tiny local uniaxial strain accidentally introduced to the sample may gives enough perturbation to fix $\bm{\hat{n}}$.
Further experimental and theoretical studies are necessary to resolve this issue.

\clearpage

\subsection{Nematic order parameter and gap structure of the $\bm{\Delta_4}$ states}
\label{sec:Delta4_discussion}

For the $\Delta_4$ states (either $\Delta_{4x}$ or $\Delta_{4y}$ states) accompanied by in-plane gap anisotropy, one can define the multi-component nematic order parameter as $Q = (|\Psi_1|^2-|\Psi_2|^2,\ \Psi_1\Psi_2^\ast + \Psi_1^\ast\Psi_2)$, where $\bm{\Psi} = (\Psi_1, \Psi_2)$ is the multi-component SC order parameter of the $E_u$ state~\cite{Fu2014.PhysRevB.90.100509}. 
Indeed, for the $\Delta_4$ states, which is expressed as $\bm{\Psi} \propto (\cos\theta,\ \sin\theta)$ with $\theta$ as a parameter, $Q \propto (\cos 2\theta,\ \sin 2\theta)$ exhibits certainly a non-zero value for any $\theta$.
Thus, the $\Delta_4$ states are indeed nematic superconducting states with well-defined and finite nematic order parameter.

Originally, the nematic $\Delta_4$ state is thought to have point nodes along an in-plane direction~\cite{Fu2010.PhysRevLett.105.097001}.
However, because these point nodes are {\em not} generally protected by symmetry, the nodes are easily gapped out by additional contributions, such as terms of higher order in $k$ in the normal-state Hamiltonian, as discussed in Ref.~\cite{Fu2014.PhysRevB.90.100509}.
Thus, the $\Delta_4$ state is in most cases fully gapped, as represented by the $\Delta_{4y}$ state (see Fig.~\ref{fig1}D).
Exceptions are when the nodes are protected by an additional crystalline symmetry, such as the mirror symmetry with respect to the $yz$ plane as in the case for the $\Delta_{4x}$ state.

In most cases, a fully gapped state is energetically more stable than a nodal state, unless the pairing interaction strongly favors the latter, because the system can gain higher condensation energy for a fully gapped state.
Thus, our conclusion on the realization of the (probably fully gapped) $\Delta_{4y}$ state agrees with this energetical expectation. 
We also note that a fully gapped state is again consistent with the previous study on the temperature dependence of the specific heat~\cite{Kriener2011.PhysRevLett.106.127004}.

There is a possibility that the Fermi surface of superconducting \cbs\ has a quasi-two-dimensional (Q2D) cylindrical shape, rather than a three-dimensional (3D) ellipsoidal shape~\cite{Lahoud2013.PhysRevB.88.195107}. 
As discussed in \ref{sec:cone}, it is difficult to distinguish the two possibilities only from our experimental and theoretical data.
Nevertheless, the nematic $\Delta_{4x}$ or $\Delta_{4y}$ SC states can be realized for both Q2D and 3D cases.
The only difference is that the point gap minima (or nodes) in the 3D case will be replaced with line gap minima in the extreme 2D limit~\cite{Hashimoto2014.SupercondSciTechnol.27.104002}. 
Thus, our conclusion on the realization of the nematic SC state holds for both the 3D and Q2D cases.

\clearpage

\subsection{Determination of $\bm{\Hcc}$}
\label{sec:Hc2}

\begin{figure}[htbp]
\begin{center}
\includegraphics[width=16cm]{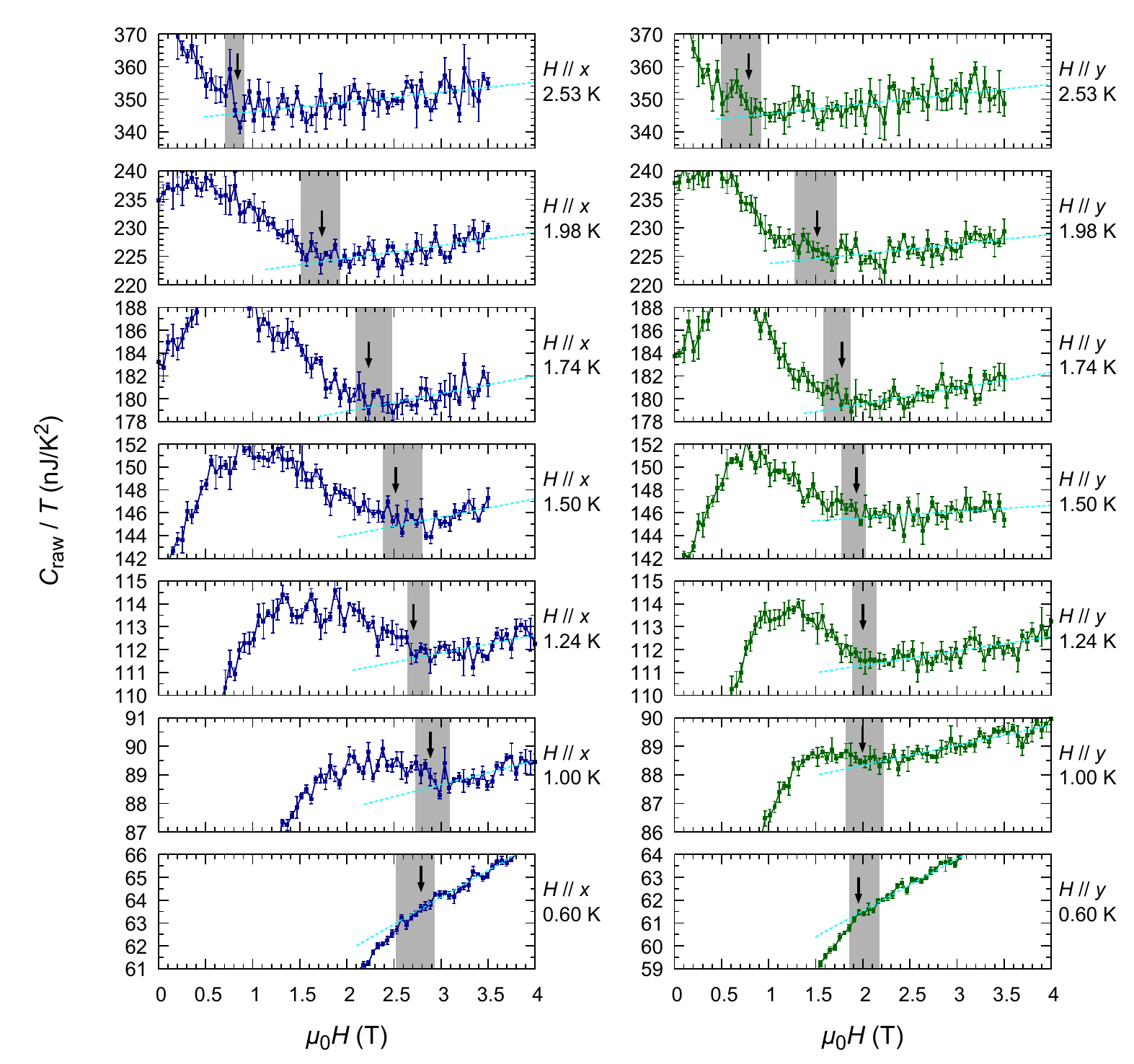}
\end{center}
\caption{
{\bf Magnetic-field dependence of the heat capacity of \cbsbm\ at various temperatures.} 
The field-dependence of $\C\subm{raw}/T$ (raw data: including phonon and background contributions) for $\bm{H}\parallel \bm{x}$ and $\bm{H}\parallel \bm{y}$ are plotted in the left and right columns, respectively.
The arrows indicate $\Hcc$ (defined as the onset of the deviation from the linear field dependence in the N state) plotted in Fig.~\ref{fig3}E, with the shaded regions presenting error bars in $\Hcc$. 
The broken lines present results of linear fittings to the data above $\Hcc$.
\label{fig:H-sweep}
}
\end{figure}

We determined the upper critical field $\Hcc$ as the onset field of the deviation from linear field dependence of $\C/T$ in the normal state above $\Hcc$.
In Fig.~\ref{fig:H-sweep}, we present the raw data of the heat capacity $\C\subm{raw}(H)/T$ (including phonon and background contributions) at various temperatures for $\bm{H}\parallel \bm{x}$ and $\bm{H}\parallel \bm{y}$.
In this figure, $\Hcc$ is marked by arrows and its error range is indicated with shaded rectangles.
Although the anomaly at $\Hcc$ is rather vague especially at low temperatures, we can define $\Hcc$ consistently throughout the investigated temperature range. 
The obtained $\Hcc$ is plotted in Fig.~\ref{fig3}E (in the main text) as a function of temperature.
By extrapolating $\Hcc(T)$ to $T=0$~K, we obtain the zero-temperature values $\mu_0\Hcc(0)\simeq 2.9$~T for the $\bm{x}$ direction and $\mu_0\Hcc(0)\simeq 2.0$~T for the $\bm{y}$ direction.

\begin{figure}[htbp]
\begin{center}
\includegraphics[width=16cm]{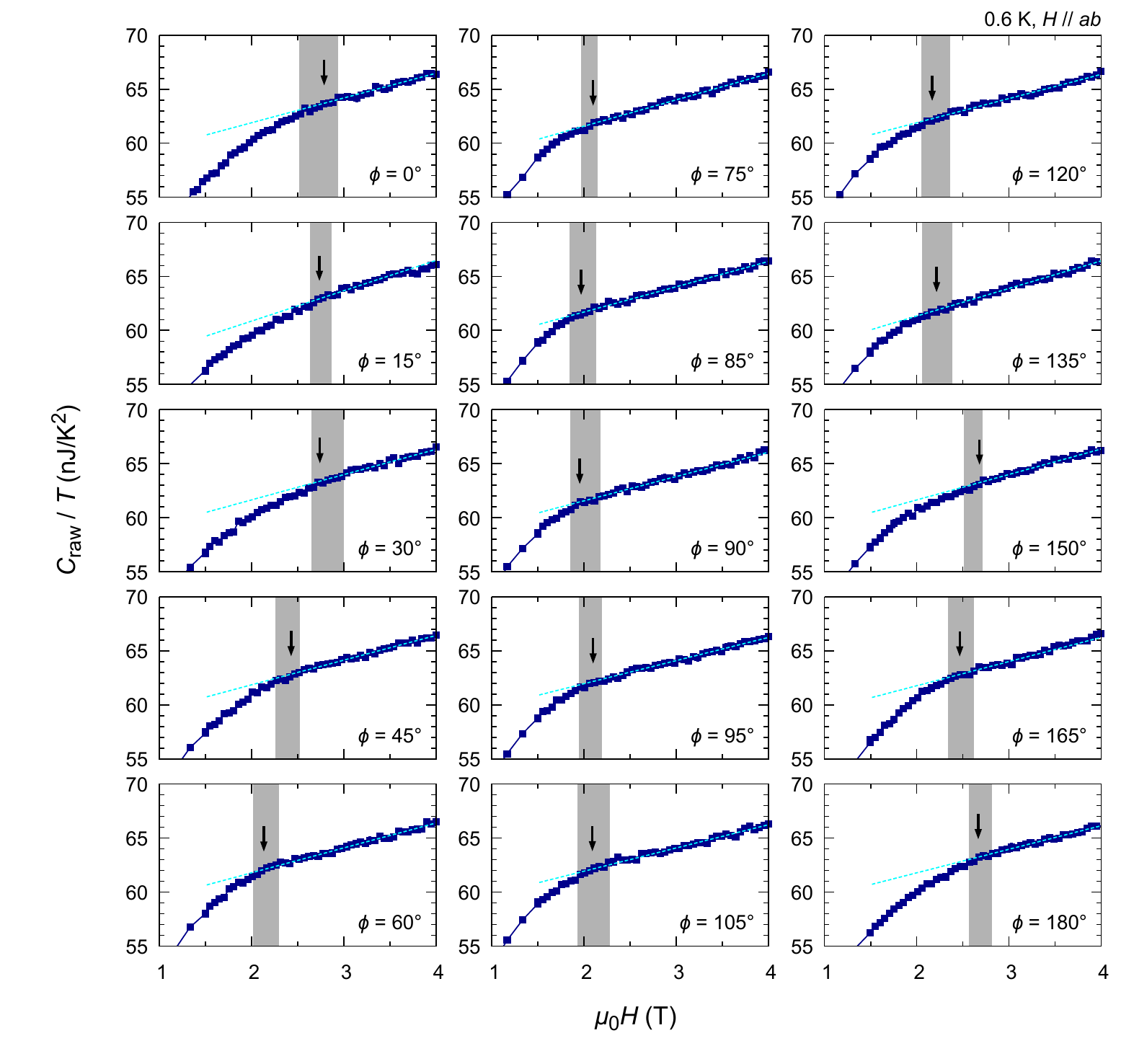}
\end{center}
\caption{
{\bf Magnetic-field dependence of the heat capacity of \cbsbm\ at 0.6~K for various in-plane field directions.} 
The field-dependence of $\C\subm{raw}/T$ (raw data: including phonon and background contributions) for various in-plane filed directions $\phi$ are plotted.
The arrows indicate $\Hcc$ (defined as the onset of the deviation from the linear field dependence in the N state) plotted in Fig.~\ref{fig2}C, with the shaded regions presenting error bars in $\Hcc$. 
The broken lines present results of linear fittings to the data above $\Hcc$.
\label{fig:H-sweep_phi-dep}
}
\end{figure}

In Fig.~\ref{fig:H-sweep_phi-dep}, we show $\C\subm{raw}(H)/T$ at 0.6~K for various in-plane field directions.
From these data, we deduce the $\phi$ dependence of $\Hcc$, which is presented in Fig.~\ref{fig2}D in the main text.

%

%

\subsection{Other superconductors exhibiting bulk RSB}
\label{sec:other_SCs}

The bulk RSB in the SC state was reported in the thermal conductivity measured in the B phase of UPt\sub{3} (trigonal structure; space group $P\bar{3}m1$, $D^3_{3d}$, No.~164~\cite{Walko2001.PhysRevB.63.054522})~\cite{Machida2012.PhysRevLett.108.157002}.
However, in this system, the rotational symmetry must have been already broken in the normal state, as evidenced by the multiple superconducting transitions at zero field~\cite{Fisher1989.PhysRevLett.62.1411}.
This normal state symmetry breaking is probably due to the underlying antiferromagnetic (short-range) order~\cite{Aeppli1988.PhysRevLett.60.615}.
Similarly, two-fold oscillation in the field-angle dependence of the thermal conductivity of PrOs\sub{4}Sb\sub{12} (cubic structure; space group $Im\bar{3}$, $T^5_h$, No.~204~\cite{Braun1980.JLessCommonMet.72.147}) was reported~\cite{Izawa2003.PhysRevLett.90.117001}.
Nevertheless, four-fold rotation is actually {\em not} a symmetry operation of the space group $Im\bar{3}$ in spite of its cubic structure: i.e. the four-fold rotational symmetry is already broken in the crystal lattice.

Therefore, the reported RSB of superconductivity in these compounds cannot be termed as ``spontaneous'' and hence ``nematic''.
We also emphasize that the RSB was {\em not} observed in the specific heat for both materials~\cite{Sakakibara2007.JPhysSocJpn.76.051004.review,Kittaka2013.JPhysSocJpn.82.024707}.

To conclude, \cbs\ is the first superconductor that clearly exhibits {\em spontaneous} RSB in a {\em thermodynamic} quantity.

\clearpage


\begin{thebibliography}{10}

\bibitem{Sigrist1991.RevModPhys.63.239}
M.~Sigrist, K.~Ueda, {\it Rev. Mod. Phys.\/} {\bf 63}, 239 (1991).

\bibitem{Tsuei2000.RevModPhys.72.969}
C.~C. Tsuei, J.~R. Kirtley, {\it Rev. Mod. Phys.\/} {\bf 72}, 969 (2000).

\bibitem{Nelson2004.Science.306.1151}
K.~D. Nelson, Z.~Q. Mao, Y.~Maeno, Y.~Liu, {\it Science\/} {\bf 306}, 1151
  (2004).

\bibitem{Fu2014.PhysRevB.90.100509}
L.~Fu, {\it Phys. Rev. B\/} {\bf 90}, 100509(R) (2014).

\bibitem{Ando2002.PhysRevLett.88.137005}
Y.~Ando, K.~Segawa, S.~Komiya, A.~N. Lavrov, {\it Phys. Rev. Lett.\/} {\bf 88},
  137005 (2002).

\bibitem{Borzi2007.Science.315.214}
R.~A. Borzi {\it et~al.\/}, {\it Science\/} {\bf 315}, 214 (2007).

\bibitem{Kasahara2012.Nature.486.382}
S.~Kasahara {\it et~al.\/}, {\it Nature\/} {\bf 486}, 382 (2012).

\bibitem{Okazaki2011.Science.331.439}
R.~Okazaki {\it et~al.\/}, {\it Science\/} {\bf 331}, 439 (2011).

\bibitem{Hor2010.PhysRevLett.104.057001}
Y.~S. Hor {\it et~al.\/}, {\it Phys. Rev. Lett.\/} {\bf 104}, 057001 (2010).

\bibitem{Fu2010.PhysRevLett.105.097001}
L.~Fu, E.~Berg, {\it Phys. Rev. Lett.\/} {\bf 105}, 097001 (2010).

\bibitem{Ando2015.AnnuRevCondensMatterPhys.6.361}
Y.~Ando, L.~Fu, {\it Annu. Rev. Condens. Matter Phys.\/} {\bf 6}, 361 (2015).

\bibitem{Sasaki2015.PhysicaC.514.206}
S.~Sasaki, T.~Mizushima, {\it Physica C\/} {\bf 514}, 206 (2015).

\bibitem{Nagai2012.PhysRevB.86.094507}
Y.~Nagai, H.~Nakamura, M.~Machida, {\it Phys. Rev. B\/} {\bf 86}, 094507
  (2012).

\bibitem{Sasaki2011.PhysRevLett.107.217001}
S.~Sasaki {\it et~al.\/}, {\it Phys. Rev. Lett.\/} {\bf 107}, 217001 (2011).

\bibitem{Levy2013.PhysRevLett.110.117001}
N.~Levy {\it et~al.\/}, {\it Phys. Rev. Lett.\/} {\bf 110}, 117001 (2013).

\bibitem{Mizushima2014.PhysRevB.90.184516}
T.~Mizushima, A.~Yamakage, M.~Sato, Y.~Tanaka, {\it Phys. Rev. B\/} {\bf 90},
  184516 (2014).

\bibitem{Lahoud2013.PhysRevB.88.195107}
E.~Lahoud {\it et~al.\/}, {\it Phys. Rev. B\/} {\bf 88}, 195107 (2013).

\bibitem{Zheng2015.unpublished}
K. Matano, M. Kriener, K. Segawa, Y. Ando, and Guo-qing Zheng;
http://arXiv.org/abs/1512.07086 (2015).

\bibitem{MaterialMethods}
Materials and Methods are available as Supplementary Mmaterials on Science
  Online.

\bibitem{Vekhter1999}
I.~Vekhter, P.~J. Hirschfeld, J.~P. Carbotte, E.~J. Nicol, {\it Phys. Rev. B\/}
  {\bf 59}, R9023 (1999).

\bibitem{Sakakibara2007.JPhysSocJpn.76.051004.review}
T.~Sakakibara {\it et~al.\/}, {\it J. Phys. Soc. Jpn.\/} {\bf 76}, 051004
  (2007).

\bibitem{VorontsovA2006.PhysRevLett.96.237001}
A.~Vorontsov, I.~Vekhter, {\it Phys. Rev. Lett.\/} {\bf 96}, 237001 (2006).

\bibitem{An2010.PhysRevLett.104.037002}
K.~An {\it et~al.\/}, {\it Phys. Rev. Lett.\/} {\bf 104}, 037002 (2010).

\newcounter{firstbib}
\setcounter{firstbib}{\value{enumiv}}
%
\end{thebibliography}

\begin{thebibliography}{10}
\setcounter{enumiv}{\value{firstbib}}
\bibitem{Kriener2011.PhysRevLett.106.127004}
M.~Kriener, K.~Segawa, Z.~Ren, S.~Sasaki, Y.~Ando, {\it Phys. Rev. Lett.\/}
  {\bf 106}, 127004 (2011).

\bibitem{Kriener2011.PhysRevB.84.054513}
M.~Kriener {\it et~al.\/}, {\it Phys. Rev. B\/} {\bf 84}, 054513 (2011).

\bibitem{Yonezawa2013.PhysRevLett.110.077003}
S.~Yonezawa, T.~Kajikawa, Y.~Maeno, {\it Phys. Rev. Lett.\/} {\bf 110}, 077003
  (2013).

\bibitem{Sullivan1968}
P.~F. Sullivan, G.~Seidel, {\it Phys. Rev.\/} {\bf 173}, 679 (1968).

\bibitem{Velichkov1992.Cryogenics.32.285}
I.~Velichkov, {\it Cryogenics\/} {\bf 32}, 285 (1992).

\bibitem{Deguchi2004RSI}
K.~Deguchi, T.~Ishiguro, Y.~Maeno, {\it Rev. Sci. Instrum.\/} {\bf 75}, 1188
  (2004).

\bibitem{Nagai2011.PhysRevB.83.104523}
Y.~Nagai, H.~Nakamura, M.~Machida, {\it Phys. Rev. B\/} {\bf 83}, 104523
  (2011).

\bibitem{Nagai2014.JPhysSocJpn.83.063705}
Y.~Nagai, {\it J. Phys. Soc. Jpn.\/} {\bf 83}, 063705 (2014).

\bibitem{Hashimoto2014.SupercondSciTechnol.27.104002}
T.~Hashimoto, K.~Yada, A.~Yamakage, M.~Sato, Y.~Tanaka, {\it Supercond. Sci.
  Technol.\/} {\bf 27}, 104002 (2014).

\bibitem{Walko2001.PhysRevB.63.054522}
D.~A. Walko {\it et~al.\/}, {\it Phys. Rev. B\/} {\bf 63}, 054522 (2001).

\bibitem{Machida2012.PhysRevLett.108.157002}
Y.~Machida {\it et~al.\/}, {\it Phys. Rev. Lett.\/} {\bf 108}, 157002 (2012).

\bibitem{Fisher1989.PhysRevLett.62.1411}
R.~A. Fisher {\it et~al.\/}, {\it Phys. Rev. Lett.\/} {\bf 62}, 1411 (1989).

\bibitem{Aeppli1988.PhysRevLett.60.615}
G.~Aeppli {\it et~al.\/}, {\it Phys. Rev. Lett.\/} {\bf 60}, 615 (1988).

\bibitem{Braun1980.JLessCommonMet.72.147}
D.~Braun, W.~Jeitschko, {\it J. Less-Common Met.\/} {\bf 72}, 147 (1980).

\bibitem{Izawa2003.PhysRevLett.90.117001}
K.~Izawa {\it et~al.\/}, {\it Phys. Rev. Lett.\/} {\bf 90}, 117001 (2003).

\bibitem{Kittaka2013.JPhysSocJpn.82.024707}
S.~Kittaka {\it et~al.\/}, {\it J. Phys. Soc. Jpn.\/} {\bf 82}, 024707 (2013).

\end{thebibliography}
\end{document}